\documentclass[aps,prd,preprintnumbers,twocolumn,groupedaddress,nofootinbib,showpacs]{revtex4-1}
\usepackage{graphicx}
\usepackage{amsmath}
\usepackage{amssymb}
\usepackage{mathtools}
\usepackage{subfigure}
\usepackage{float} 
\usepackage{color}

\def\lsim{\mathrel{\raise.3ex\hbox{$<$\kern-.75em\lower1ex\hbox{$\sim$}}}}
\def\gsim{\mathrel{\raise.3ex\hbox{$>$\kern-.75em\lower1ex\hbox{$\sim$}}}}

\begin{document}

\title{Stringent Constraints On The Dark Matter Annihilation Cross Section From Subhalo Searches With The Fermi Gamma-Ray Space Telescope}  
\author{Asher Berlin$^{1}$}
\author{Dan Hooper$^{2,3}$}
\affiliation{$^1$Department of Physics, University of Chicago, 5720 S Ellis Ave., Chicago, IL 60637}
\affiliation{$^2$Department of Astronomy and Astrophysics, University of Chicago, 5640 S Ellis Ave., Chicago, IL 60637}
\affiliation{$^3$Center for Particle Astrophysics, Fermi National Accelerator Laboratory, Batavia, IL 60510}
\date{\today}

\begin{abstract}

The dark matter halo of the Milky Way is predicted to contain a very large number of smaller subhalos. As a result of the dark matter annihilations taking place within such objects, the most nearby and massive subhalos could appear as point-like or spatially extended gamma-ray sources, without observable counterparts at other wavelengths. In this paper, we use the results of the Aquarius simulation to predict the distribution of nearby subhalos, and compare this to the characteristics of the unidentified gamma-ray sources observed by the Fermi Gamma-Ray Space Telescope. Focusing on the brightest high latitude sources, we use this comparison to derive limits on the dark matter annihilation cross section. For dark matter particles lighter than $\sim$200 GeV, the resulting limits are the strongest obtained to date, being modestly more stringent than those derived from observations of dwarf galaxies or the Galactic Center. We also derive independent limits based on the lack of unidentified gamma-ray sources with discernible spatial extension, but these limits are a factor of $\sim$2-10 weaker than those based on point-like subhalos. Lastly, we note that four of the ten brightest high-latitude sources exhibit a similar spectral shape, consistent with 30-60 GeV dark matter particles annihilating to $b\bar{b}$ with an annihilation cross section on the order of $\sigma v \sim (5-10)\times 10^{-27}$ cm$^3$/s, or 8-10 GeV dark matter particles annihilating to $\tau^+ \tau^-$ with $\sigma v \sim (2.0-2.5)\times 10^{-27}$ cm$^3$/s.

\end{abstract}

\pacs{95.35.+d, 95.95.Pw, 07.85.-m, FERMILAB-PUB-13-350-A}
\maketitle

\section{Introduction}

Numerical simulations of cold, collisionless dark matter particles predict patterns of large scale structure that are in excellent agreement with observations, from galaxy-scales to that of the largest superclusters, voids, and filaments. Such simulations demonstrate that dark matter halos form hierarchically, with dark matter particles first collapsing into small gravitationally bound systems, which go on to form more massive halos through a sequence of repeated mergers~\cite{White:1991mr}. A consequence of this process is that individual dark matter halos contain very large numbers of smaller subhalos.  For dark matter candidates with weak-scale masses and interactions, subhalos as small as roughly $\sim$$10^{-6}\, M_{\odot}$ are predicted to form~\cite{Profumo:2006bv}. In this case, the dark matter halo of the Milky Way is expected to contain $\sim$$10^{16}$ subhalos within its virial radius, mostly consisting of very low mass structures, but also extending up to the largest observed satellites, such as those containing dwarf spheroidal galaxies, and the $\sim$$10^{10} M_{\odot}$ Large Magellanic Cloud. 

The Fermi Gamma-Ray Space Telescope (Fermi), as well as ground-based atmospheric cherenkov telescopes, are capable of placing constraints on the nature of dark matter by searching for their annihilation products. Currently, the strongest constraints on the dark matter's annihilation cross section have been derived from gamma-ray observations of dwarf galaxies~\cite{GeringerSameth:2011iw,Ackermann:2011wa,Abramowski:2010aa} and the Galactic Center~\cite{Hooper:2012sr}  (for annihilations to $e^+ e^-$ or $\mu^+ \mu^-$, cosmic ray measurements from AMS also provide strong constraints~\cite{Bergstrom:2013jra}). If dark matter particles annihilate with a cross section not very far below the maximum value allowed by these constraints, it may be possible to observe gamma-rays from a number of nearby dark matter subhalos~\cite{Kuhlen:2008aw,Pieri:2007ir}. Such subhalos would appear as point-like or somewhat extended gamma-ray sources, without associated emission at other wavelengths.  

The prospects for observing gamma-rays from dark matter subhalos depend on their local number density and density profiles, as well as on the dark matter's particle's mass, annihilation cross section, and dominant annihilation channels.  Fortunately, high-resolution N-body simulations provide us with a relatively detailed description of the subhalo population predicted to inhabit a Milky Way-like halo. Of particular utility in this respect are the results of the Aquarius and Via Lactea II simulations.  The Aquarius Project identified approximately 300,000 subhalos within a simulated Milky Way-like system, resolving objects with masses as small as $3.24\times 10^4 M_{\odot}$~\cite{Springel:2008cc}.  The results of the Via Lactea II simulation also support this picture, although with somewhat lower resolution~\cite{Diemand:2008in}. These and other simulations provide a quantitative confirmation of the long held expectation that halos of cold, collisionless dark matter particles will contain large populations of compact subhalos.

In this paper, we revisit Fermi's ability to potentially observe annihilation products from dark matter subhalos, and to use the lack of subhalo candidate gamma-ray sources to place upper limits on the dark matter's annihilation cross section. In doing so, we expand on previous work~\cite{Belikov:2011pu,Buckley:2010vg,Zechlin:2011wa,Mirabal:2012em,Mirabal:2010ny,Zechlin:2012by} in a number of ways. Firstly, we consider not only subhalos which would appear as point-like sources to Fermi, but also derive limits based on searches for spatially extended sources, as applicable to the case of particularly large or nearby subhalos. We also update the list of Fermi's currently unidentified gamma-ray sources, removing a number of recently identified active galactic nuclei and pulsars, and perform a detailed study of the spectra and luminosities of these sources. Taking this information together, we find that the greatest sensitivity to dark matter annihilations can be extracted from the observations of Fermi's brightest unidentified, high-latitude ($|b| > 30^{\circ}$) gamma-ray sources. In particular, by studying the spectrum of the 10 such sources with fluxes greater than $10^{-9}$ photons per cm$^2$ s between 1-100 GeV, we are able to place constraints on the dark matter's annihilation cross section which are more stringent than those derived from observations of dwarf galaxies or the Galactic Center. For annihilations to $b\bar{b}$ with a cross section of $\sigma v =  3\times 10^{-26}$ cm$^3$/s, for example, we exclude dark matter masses below 100 GeV. Furthermore, as these results are based on the data collected over the first two years of Fermi's mission (as presented in the Fermi-LAT Second Source Catalog~\cite{Fermi-LAT:2011iqa}), an updated catalog with more precise spectral information would likely make it possible to improve upon these constraints significantly. Additional information from ground based gamma-ray telescopes, as well as multi-wavelength observations, could also be used to exclude many sources as subhalo candidates, potentially strengthening these limits further.

The remainder of this article is structured as follows. In Sec.~\ref{theory}, we describe our calculations for predicting the number of dark matter subhalos detectable by Fermi. In Sec.~\ref{sources}, we discuss the characteristics of Fermi's unidentified gamma-ray sources. In Sec.~\ref{constraints}, we make use of this information to derive upper limits on the dark matter's annihilation cross section. In Sec.~\ref{discussion}, we discuss the uncertainties involved in our calculation and consider the implications of our results. Lastly, in Sec.~\ref{summary}, we summarize our results and conclusions.


\section{Dark Matter Subhalos in the Milky Way}
\label{theory}

Throughout this study, we base our analysis on the results of the Aquarius Project, which has provided the highest resolution simulations to date of the dark matter subhalo populations found within the halos of Milky Way-like galaxies. In their simulations of six Milky Way-like systems, Aquarius resolved hundreds of thousands of subhalos, which were found to follow a mass distribution of the form $dN/dM \propto M^{-1.9}$~\cite{Springel:2008cc}. The shape of this mass function exhibits no discernible dependance on the location within the parent halo. And although 17.7\% of the total dark matter mass in the system was found to be in gravitationally bound objects, subhalos made up a significantly smaller fraction of the mass in the equivalent of the local region of our galaxy (at $\sim8.5$ kpc from the Galactic Center).\footnote{This percentage was calculated by extrapolating the mass distribution of subhalos down to the Earth-mass scale. Considering only those subhalos large enough to be resolved by Aquarius ($M > 3.24\times 10^{4} M_{\odot}$), the fraction of the total mass in subhalos is a slightly more modest 13.2\%. For a distribution that follows $dN/dM \propto M^{-1.9}$, most of the mass in subhalos is found in objects above the resolution of Aquarius, allowing our results to be only mildly sensitive to the presence of any lower mass subhalos.} This is understood to be the consequence of tidal effects; subhalos traveling in orbits passing through the inner volume of the host halo encounter other subhalos more frequently, and thus generally experience a much greater degree of mass loss than those subhalos located further from the Galactic Center. 


\begin{figure*}
\mbox{\includegraphics[width=0.49\textwidth,clip]{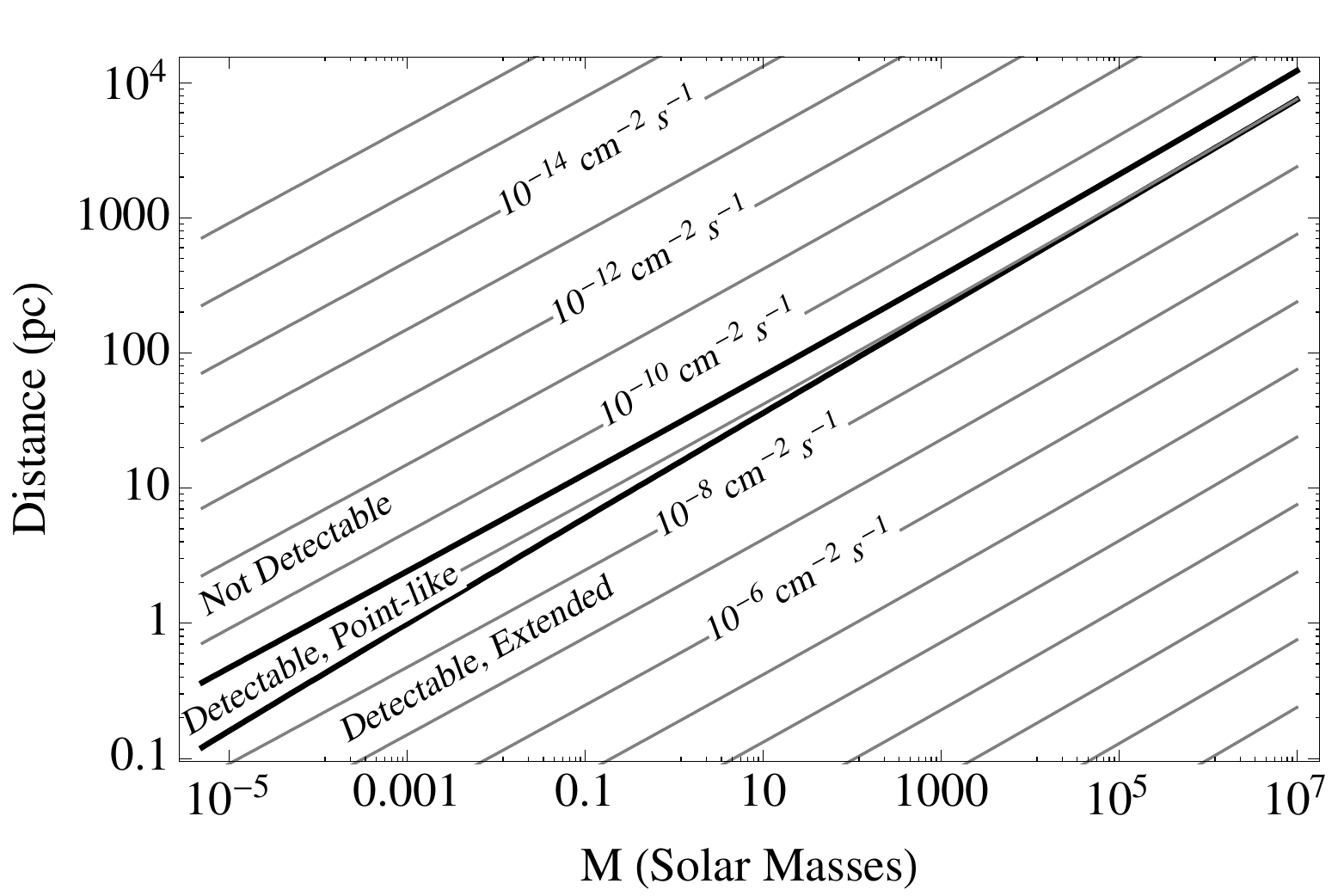}} 
\mbox{\includegraphics[width=0.49\textwidth,clip]{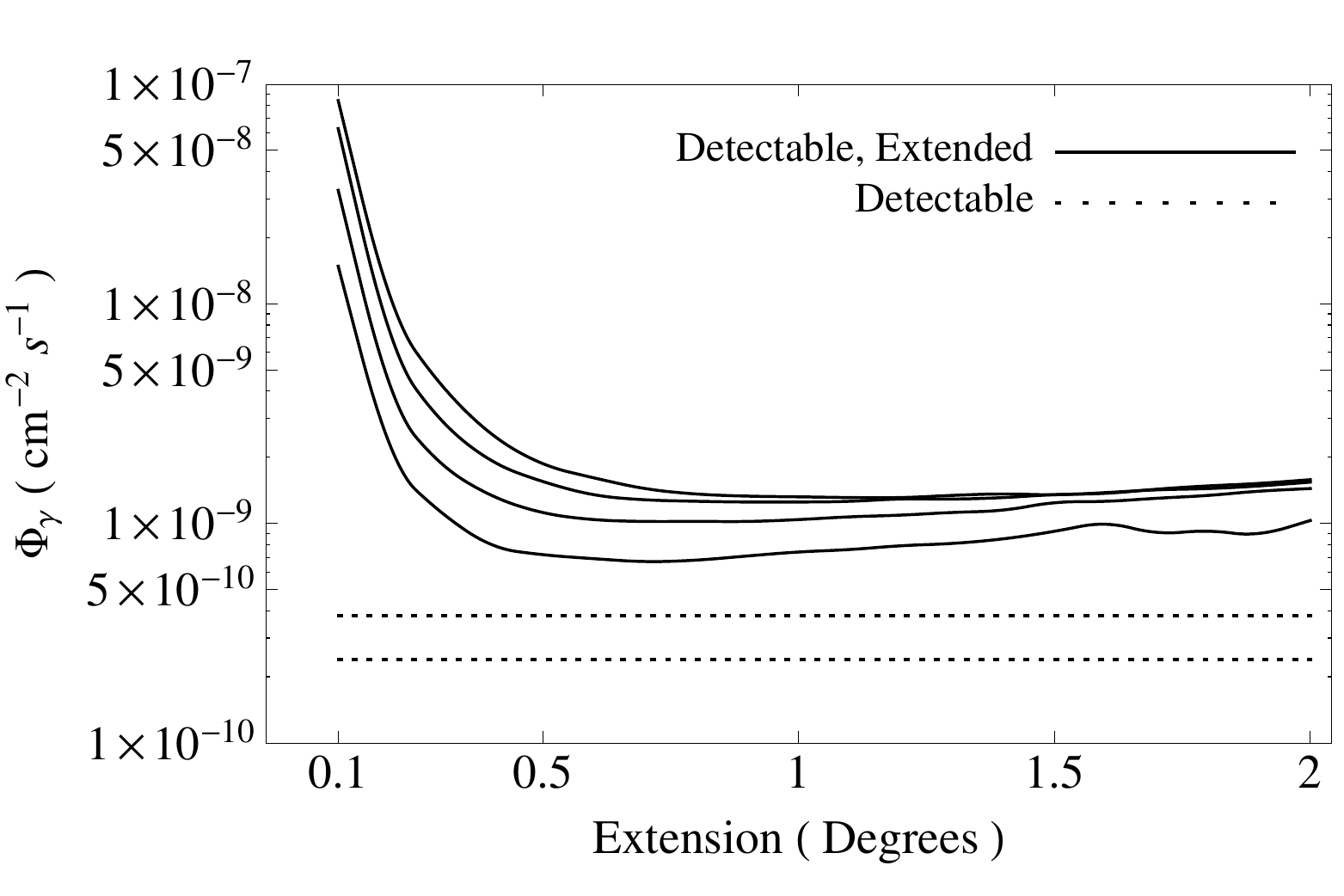}} 
\caption{Left: Regions of the mass-distance plane for which a given subhalo will be detectable by Fermi, as either a point-like source or as a source with discernible spatial extension, for the case of a 100 GeV dark matter particle with an annihilation cross section of $\sigma v = 3\times 10^{-26}$ cm$^3$/s to $b\bar{b}$. Also shown are contours of constant gamma-ray flux (1-100 GeV). Right: The flux threshold (1-100 GeV) for Fermi to detect a source outside of the Galactic Plane ($|b| > 10^{\circ}$), for spectral indices between 1.5 and 3.0 (bracketed by the two dotted lines)~\cite{Fermi-LAT:2011iqa}, and to detect spatial extension from such a source, as a function of its spatial extension (solid lines, for spectral indices of 3.0, 2.5, 2.0, and 1.5, from top-to-bottom)~\cite{Lande:2012xn}. See text for details.}
\label{detect}
\end{figure*}

Based on the results of the Aquarius simulation (the subhalo distribution shown in Fig.~11 of Ref.~\cite{Springel:2008cc}, combined with the total number of resolved subhalos for halo Aq-A-1, as reported in Table 2 of the same paper), we adopt the following distribution for subhalos in the local region of the Milky Way:
\begin{equation}
\frac{dN}{dM dV} = 260 \,\, {\rm kpc}^{-3} \, M_{\odot}^{-1} \times \bigg(\frac{M}{M_{\odot}}\bigg)^{-1.9}.  
\label{norm}
\end{equation}
Integrating this expression down to $M = 3.24 \times 10^4 \, M_{\odot}$ (the resolution of Aquarius) yields a local mass density in subhalos of 7350 $M_{\odot}/$kpc$^3$ (0.000292 GeV/cm$^3$), corresponding to approximately 0.073\% of the overall local dark matter density (in good agreement with Fig.~12 of Ref.~\cite{Springel:2008cc}). 

For each individual subhalo, we begin by considering an initial (at the time of infall, prior to tidal effects) dark matter distribution which follows an Einasto profile:
\begin{equation}
\rho(r) \propto \exp\bigg[-\frac{2}{\alpha}\bigg(\frac{r^{\alpha}}{r^{\alpha}_{-2}}-1\bigg)\bigg],
\end{equation}
where $r$ is the distance to the center of the subhalo, $\alpha=0.16$, and $r_{-2}$ is the radius at which $\rho(r) \propto r^{-2}$ (or equivalently, the radius at which $\frac{d\rho}{dr}\frac{r}{\rho}=-2$), analogous to the scale radius of an NFW profile. For subhalos located in the outer volume of the Milky Way's halo (near the virial radius), the initial dark matter distribution will remain largely intact. For the nearby subhalos that we are most interested in, however, tidal stripping will remove the vast majority of the total mass from each subhalo. In particular, by comparing the fractions of mass in subhalos in the local volume and near the virial radius of the Milky Way, as found in Ref.~\cite{Springel:2008cc}, it can be seen that approximately 99.5\% of the local mass in subhalos has been lost to tidal effects. We adopt this fractional mass loss throughout our analysis, assuming that each nearby subhalo has lost the outermost 99.5\% of its initial mass, leaving behind only the most centrally concentrated volume of the Einasto profile described above. For the initial concentration of each subhalo (prior to tidal effects), defined as the ratio of the virial and scale radii, $c \equiv r_{\rm vir}/r_{-2}$, we adopt the values presented in Ref.~\cite{MunozCuartas:2010ig}, with subhalo-to-subhalo variations modeled by a log-normal distribution with a dispersion of $\sigma_c=0.24$~\cite{Bullock:1999he}.

The differential gamma-ray spectrum per solid angle from dark matter annihilations within an individual subhalo is given by:
\begin{align}
\label{flux1}
\Phi (E_{\gamma}, \theta) = \frac{1}{8 \pi  m_{X}^2} \sum_{i} \langle \sigma v \rangle_i \frac{\mathrm{d}N_{\gamma,i}}{\mathrm{d}E_\gamma}  \,  \int_{\text{l.o.s.}} \rho^2 [ r(D,l,\theta)] \,\mathrm{d}l, 
\end{align}
where $m_X$ is the mass of the dark matter particle, $\langle \sigma v \rangle_i$ is the annihilation cross section to final state $i$, and $dN_{\gamma, i}/dE_{\gamma}$ is the gamma-ray spectrum produced per annihilation to final state $i$, which we calculate using PYTHIA 8~\cite{pythia}. The integral of the density squared is performed over the line-of-sight, $D$ is the distance to the center of the subhalo, $\theta$ is the angle to the center of the subhalo, and $r(\theta,D,l)=\sqrt{D^2+l^2-2Dl\cos{\theta}}$.

The brightness and angular distribution of the gamma-rays from a subhalo depends on its mass and distance. In the left frame of Fig.~\ref{detect}, we show (for the case of a 100 GeV dark matter particle, annihilating with $\sigma v=3\times 10^{-26}$ cm$^3$/s to $b\bar{b}$) the range of distances and masses for which a subhalo will be observable by Fermi and, if observable, whether it will exhibit a discernible degree of spatial extension (in contrast to being indistinguishable from a point source).  To be detected by Fermi (and appear within the Fermi-LAT Second Source Catalog~\cite{Fermi-LAT:2011iqa}), we require that a high latitude ($|b| > 10^{\circ}$) subhalo produce a gamma-ray flux between 1-100 GeV that exceeds the threshold described in Ref.~\cite{Fermi-LAT:2011iqa}. As we will show, the precise value of this threshold does not significantly impact Fermi's sensitivity to dark matter annihilations in subhalos, as the constraints are dominated by the observed number of very bright sources. In generating this figure, we have adopted central values for the halo concentration~\cite{MunozCuartas:2010ig}, thus neglecting the impact of halo-to-halo variations. 

As most astrophysical sources of gamma-rays are point-like (such as pulsars and active galactic nuclei, for example), detecting spatial extension from an unidentified source could potentially be very useful in identifying it as a dark matter subhalo. In order for Fermi to detect spatial extension from a given subhalo, we require that its flux exceeds the threshold described in Ref.~\cite{Lande:2012xn}, which we reproduce in the right frame of Fig.~\ref{detect}.  This threshold is a function of the spatial extension and spectral shape of the source. By ``extension'', we (and the authors of Ref.~\cite{Lande:2012xn}) denote the angular radius of a disk of uniform luminosity per area.  To apply this threshold to the case of a dark matter subhalo (which is not a uniform disk), we consider a source to have discernible extension if less than 68\% of its photons originate from within a radius equal to 82\% of the quoted extension. Results are shown for spectral indices of $\Gamma=$~1.5, 2.0, 2.5 and 3.0, such that $dN_{\gamma}/dE_{\gamma} \propto E_{\gamma}^{-\Gamma}$. As the gamma-ray spectrum from dark matter annihilations does not take a power-law form, we apply these thresholds by choosing a spectral index that predicts the same mean photon energy (between 1-100 GeV) as the dark matter model under consideration.

\begin{figure}
\mbox{\includegraphics[width=0.49\textwidth,clip]{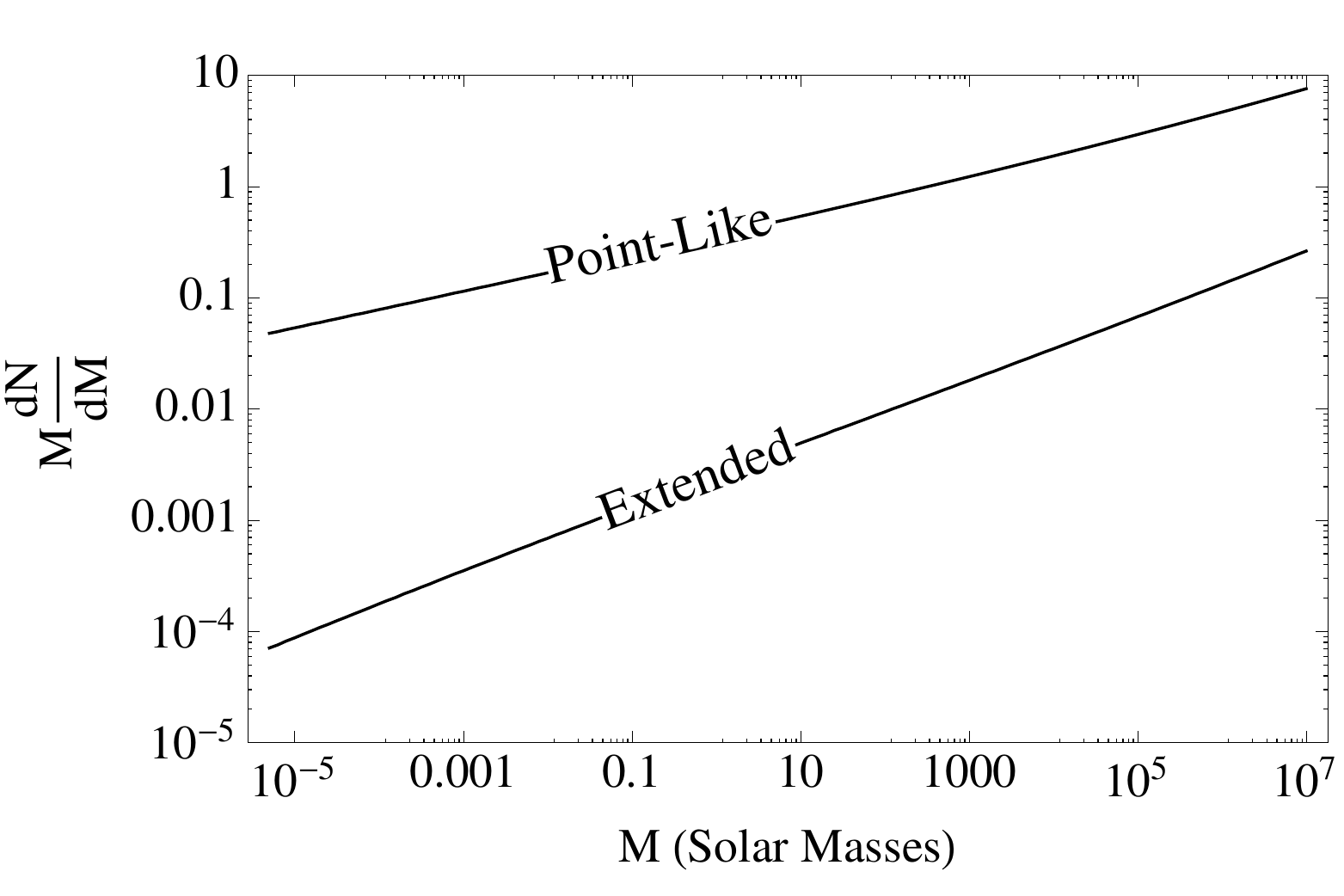}}
\caption{The distribution of detectable point-like and discernibly extended subhalos, for a 100 GeV dark matter particle annihilating to $b\bar{b}$ with a cross section of $\sigma v = 3\times 10^{-26}$ cm$^3$/s. Most of the observable subhalos are massive and appear point-like to Fermi (a few percent have a detectable degree of spatial extension).}
\label{dist}
\end{figure}

The population of detectable subhalos, both point-like and extended, is dominated by the most massive and nearby of such objects. In the calculations used to derive our constraints, we include subhalos with masses up to $10^7 M_{\odot}$. We have chosen to neglect subhalos above this mass because we expect many of the more massive subhalos to contain significant quantities of baryons (stars and/or gas) and thus will evolve to become the Milky Way's satellite galaxies (i.e.~dwarf spheroidals). From Fig.~\ref{detect}, we see that even very large subhalos are detectable by Fermi only if they are within a few kiloparsecs from the Solar System. In Fig.~\ref{dist}, we plot the distribution of point-like and discernibly extended subhalos that are detectable by Fermi, for the case of a 100 GeV dark matter particle annihilating with a cross section of $\sigma v = 3\times 10^{-26}$ cm$^3$/s to $b\bar{b}$. Most of the observable subhalos are quite massive. Only a few percent of the detectable subhalos are predicted to be discernibly extended.

We note that in some respects the results of the Aquarius simulation lead to more pessimistic predictions for subhalo searches than might be made based on Via Lactea II.  In particular, Via Lactea favors somewhat steeper density profiles ($\rho \propto r^{-1.2}$ in the innermost volumes)~\cite{Diemand:2009bm}. Also, we have conservatively neglected any enhancements to the annihilation rate that might result from dark matter substructures found within individual subhalos.

\section{Fermi's Unidentified Gamma-Ray Sources}
\label{sources}

\begin{figure}[!]
\mbox{\includegraphics[width=0.49\textwidth,clip]{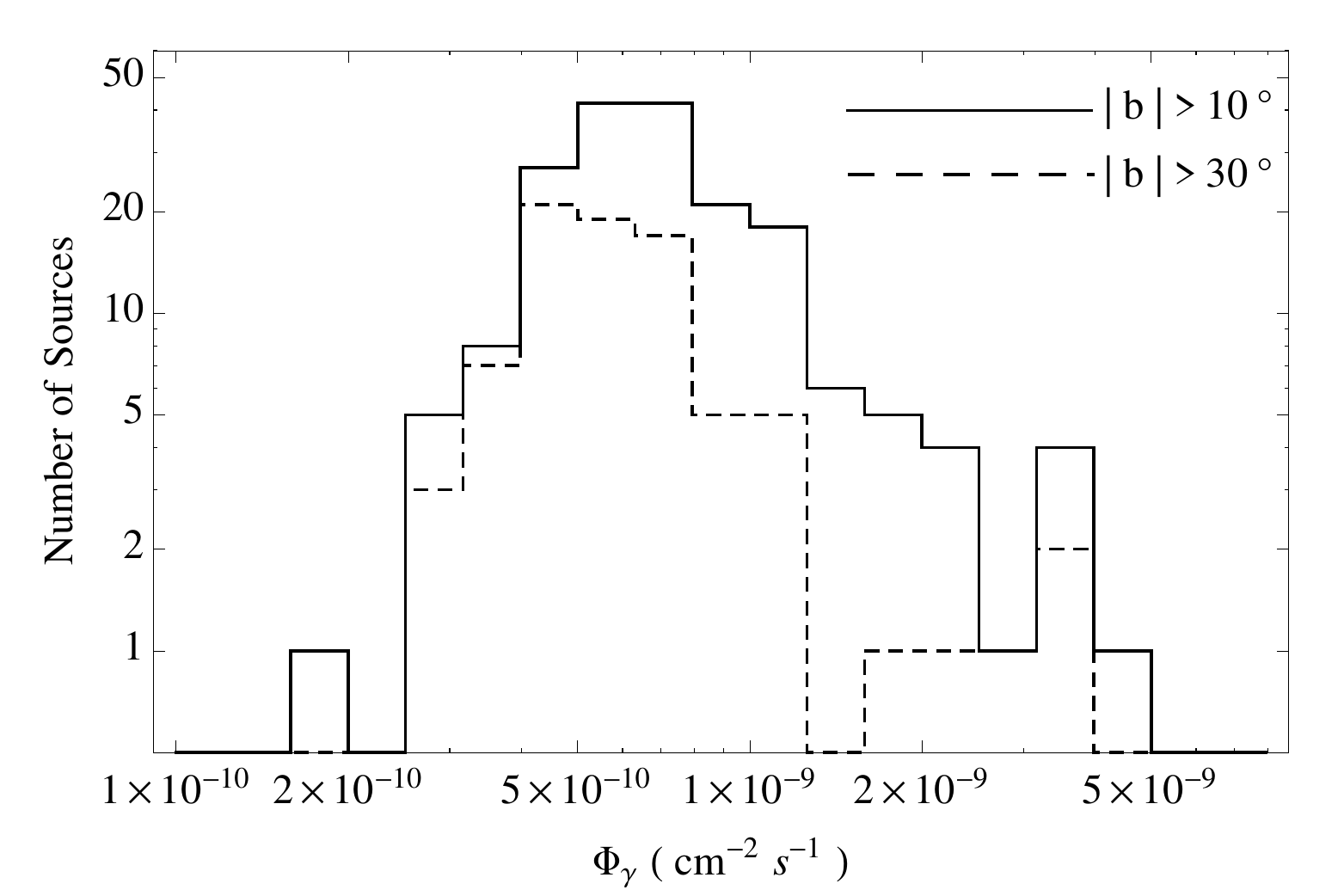}}
\caption{The flux distribution ($E_{\gamma}=1-100$ GeV) of the unidentified Fermi sources without detected variability and that are located more than $10^{\circ}$ or $30^{\circ}$ away from the Galactic Plane.}
\label{histogramall}
\end{figure}

In 2011, the Fermi Collaboration released a catalog of gamma-ray sources based on their first 24 months of data. This catalog, known as the Fermi-LAT Second Source Catalog (or the 2FGL), includes a total of 1873 gamma-ray sources, 576 of which had (at the time) not been associated with counterparts at other wavelengths. In the time since, many of these sources have been identified as either active galaxies~\cite{Massaro:2013uoa} or pulsars~\cite{pulsars}. Focusing on the subset of these unidentified sources without detected variability (variability index $<41.64$), and that are located outside of the Galactic Plane ($|b| > 10^{\circ}$), we are currently left with 185 sources to consider as potential dark matter subhalo candidates. Of these subhalo candidates, 82 of these sources are located at high galactic latitude ($|b| > 30^{\circ}$). In Fig.~\ref{histogramall}, we show the distribution of these subhalo candidate sources, as a function of flux.

\begin{table*}[t]
\centering
\begin{tabular}{|c|c|c|c|c|c|c|c|c|c|}
	\hline
Source Name&  $\Phi_{\gamma}$ (cm$^{-2}$ s$^{-1}$)    & Dec (deg)  & $b$ (deg)  & $b{\bar b}$ (GeV) &$W^+W^-$(GeV)& $ZZ$ (GeV)&$c\bar{c}$ (GeV) & $\tau^+ \tau^-$(GeV) & $\mu^+ \mu^-$(GeV) \\
	\hline
		 2FGL J2112.5-3042	&3.39e-09&-30.71	 &	-42.45&28-57&none&none&21-44&8-10&4-5\\
	 	 2FGL J1226.0+2953&1.66e-09&	29.90		&83.78 &27-66 &81-84&92-95&24-52&10-10.3&5-6\\
		 	2FGL J1511.8-0513	&1.23e-09&-5.22 &		43.12	 &34-270&81-300&92-300&29-195&5-29&3-15\\ 
	 	2FGL J1630.3+3732&1.23e-09&	37.55	&	43.26	 &25-93&81-94&92-115&21-67&8-21&3-10\\ 
			2FGL J2212.6+0702&1.01e-09&	7.05	&	-38.58 &10-24&none&none&8-10&none&none\\
			\hline	 
\end{tabular}
\caption{Fermi's brightest unidentified, non-variable, high-latitude ($|b|>30^{\circ}$) sources that exhibit a spectrum that is well-fit by at least one of the dark matter annihilation channels considered in this paper. For each source, we give its flux (between 1-100 GeV), declination, galactic latitude, and the range of masses that are well-fit to the observed spectrum for a given annihilation channel. See text for details.}
\label{tab1}
\end{table*}

\begin{table*}[t]
\centering
\begin{tabular}{|c|c|c|c|c|c|c|c|c|c|}
	\hline
Source Name&  $\Phi_{\gamma}$ (cm$^{-2}$ s$^{-1}$)    & Dec (deg)  & $b$ (deg)  & $b{\bar b}$ (GeV) &$W^+W^-$(GeV)& $ZZ$ (GeV)&$c\bar{c}$ (GeV) & $\tau^+ \tau^-$(GeV) & $\mu^+ \mu^-$(GeV) \\
	\hline
	 2FGL J1744.1-7620	&3.84e-09&-76.34	&	-22.48&24-44&none&none&17-28&6-8&none\\
	 2FGL J1539.2-3325	&2.29e-09&-33.43		&17.53&58-72&none&none&32-57&10-19&5-10\\	 
	 2FGL J0541.8-0203c&2.28e-09&	-2.06		&-16.41&8-13&none&none&5-10&3-5&1.5-2\\
		 2FGL J1722.5-0420	&1.92e-09&-4.34	 &	17.52&5-13&none&none&5-10&3-5&1.5-2.5\\
	 2FGL J1946.4-5402	&1.74e-09&-54.05		&-29.55&13-39&none&none&10-23&5-7&none\\
	 \hline
	 2FGL J0534.8-0548c&1.63e-09&	-5.81		&-19.66&26-115&81-130&92-155&23-82&5-17&3-8\\
	 2FGL J1645.7-2148c&1.52e-09&	-21.82		&15.15&5-13&none&none&5-10&none&1.5-2\\
	 2FGL J0533.9+6759&1.51e-09&	68.00		&18.19&23-73&81-88&92-101&20-57&8-13&3-5\\ 
	 	2FGL J0418.9+6636	   &1.37e-09   & 66.60	&	11.56	 &10-44&none&none&8-26&5-7&none\\ 
	 	2FGL J1805.8+0612  & 1.30e-09 &	6.21	&	12.95	 &13-69&81-85&92-98&10-53&8-10&4-6\\ 
		\hline
	 	2FGL J1544.1-2554&1.21e-09&	-25.91	&	22.61	 &8-10&none&none&none&none&none\\ 
	 	 	2FGL J0534.9-0450c &1.14e-09&	-4.84		&-19.22	 &8-57&none&none&5-40&5-8&4-5\\ 		
	 	2FGL J2017.5-1618	&1.13e-09&-16.30	&	-26.21	 &50-66&81-87&92-94&38-56&none&3-8\\
	 	2FGL J1231.3-5112	&1.10e-09&-51.21	&	11.54	 &5-10&none&none&none&4-7&none\\
	 	2FGL J1729.5-0854	&1.09e-09&-8.91	&	13.71	 &5-10&none&none&3-7&none&none\\
	\hline
			2FGL J0336.0+7504	&1.07e-09&75.08	&	15.55	 &13-57&none&none&10-39&5-10&4-5\\
	 	2FGL J1620.5-2320c&1.05e-09&	-23.34	&	18.62	 &5-8&none&none&3-7&2-5&2-2.5\\
	 	2FGL J1747.6+0324&1.03e-09&	3.40	&	15.72	 &28-125&81-140&92-165&24-84&8-17&3-10\\
	 	2FGL J1721.0+0711	&1.03e-09&7.20	&	23.36	 &8-28&none&none&5-10&3-7&1.5-3\\
	 	2FGL J1942.7-8049c&1.01e-09&	-80.82	&	-28.79	 & 5-10&none&none&5-8&2-3&none\\		
			\hline	 
	 	2FGL J1721.5-0718c&1.00e-09&	-7.30	&	16.23	 &5-10&none&none&5-8&none&none\\
		\hline	 
\end{tabular}
\caption{Fermi's brightest unidentified, non-variable, mid-latitude ($30^{\circ}>|b|>10^{\circ}$) sources that exhibit a spectrum that is well-fit by at least one of the dark matter annihilation channels considered in this paper. For each source, we give its flux (between 1-100 GeV), declination, galactic latitude, and the range of masses that are well-fit to the observed spectrum for a given annihilation channel. See text for details.}
\label{tab2}
\end{table*}


More recently, Lande {\it et al.} performed a study of the 21 spatially extended gamma-ray sources observed by Fermi~\cite{Lande:2012xn}. None of these sources, however, appears to represent a likely dark matter subhalo candidate. In particular, 17 of these 21 objects lie within 10 degrees of the Galactic Plane, and the four others are each associated with emission from known astrophysical objects: the Large and Small Magellanic Clouds, the nearby galaxy Centaurus A, and the Ophiuchus molecular cloud. Despite the lack of spatially extended subhalo candidates, the absence of of such gamma-ray sources can be used to constrain the dark matter annihilation cross section.

For each of Fermi's subhalo candidate sources, we compare the measured spectra (which, as presented in the 2FGL, consists of fluxes binned into five energy ranges (0.1-0.3 GeV, 0.3-1.0 GeV, 1.0-3.0 GeV, 3.0-10 GeV, and 10-100 GeV) to that predicted for a range of dark matter masses and annihilation channels. In order for a given source to be classified as well-fit by a given dark matter model, we require $\chi^2 < 7.77$ (over 5-1 degrees of freedom). This requirement was chosen such that 90\% of any actual dark matter subhalos will qualify, while most astrophysical sources will not. 

In Tables~\ref{tab1} and~\ref{tab2}, we list each of Fermi's bright ($\Phi_{\gamma}>10^{-9}$ cm$^{-2}$ s$^{-1}$), mid- or high-latitude ($|b|>10^{\circ}$), non-variable, unidentified sources that exhibit a spectral shape that are well-fit by at least one of the dark matter annihilation channels we have considered ($b\bar{b}$, $W^+W^-$, $ZZ$, $c\bar{c}$, $\tau^+ \tau^-$, or $\mu^+ \mu^-$). In addition to those sources included in these tables, we found that the following 2FGL sources were not well-fit by any of the dark matter annihilation channels we considered: J1902.7-7053, J2039.8-5620, J1227.7-4853, J1904.9-3720c, J0523.3-2530, J1846.6-2519, J1120.0-2204, J1704.6-0529, J1653.6-0159, J1625.2-0020, J0547.1+0020c, J1129.5+3758, and J1548.3+1453.	 While these sources could be dark matter subhalos, none of the annihilation channels considered here were found to provide a good fit to their observed spectra.

\section{Dark Matter Constraints}
\label{constraints}

\begin{figure*}[!]
\mbox{\includegraphics[width=0.49\textwidth,clip]{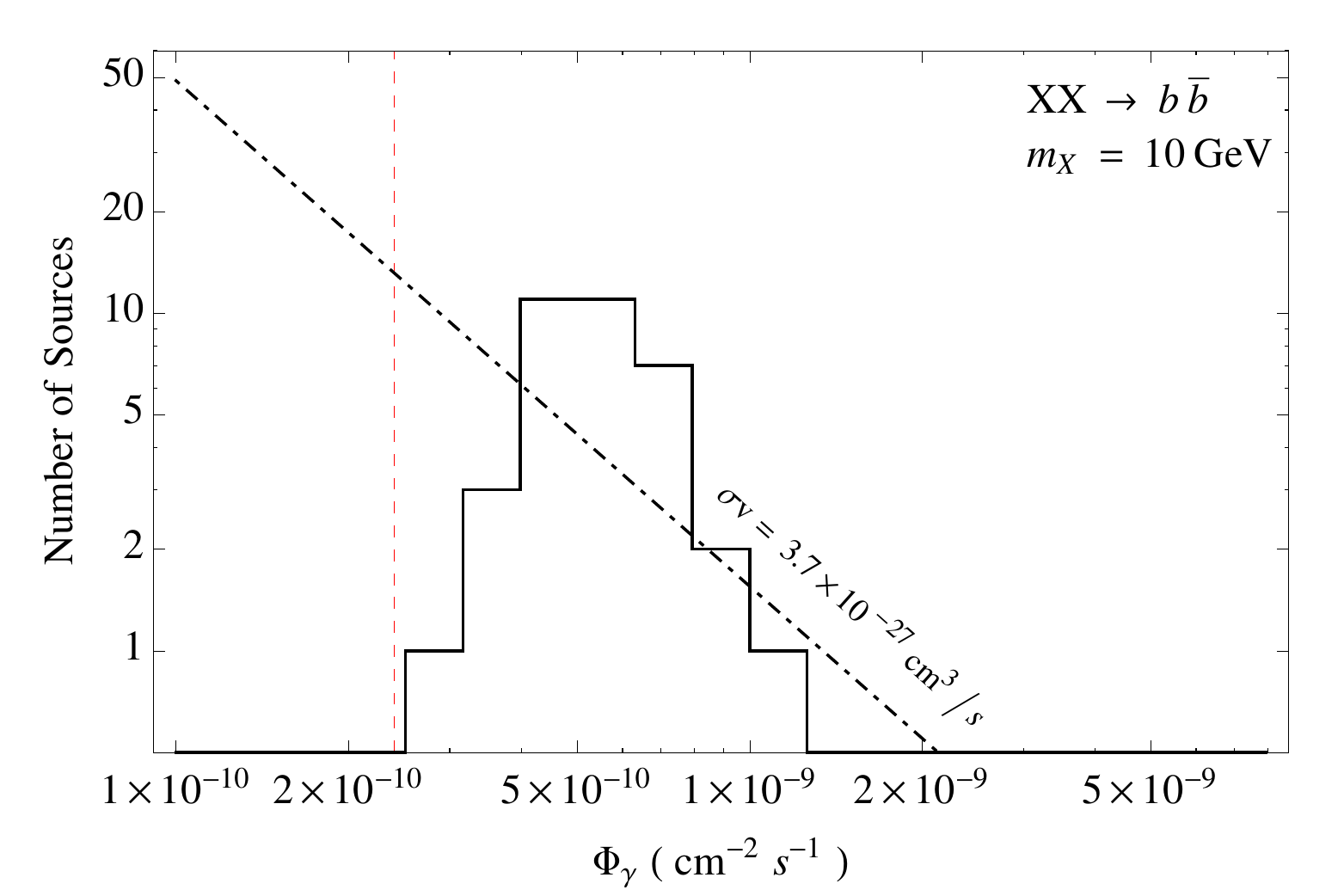}}
\mbox{\includegraphics[width=0.49\textwidth,clip]{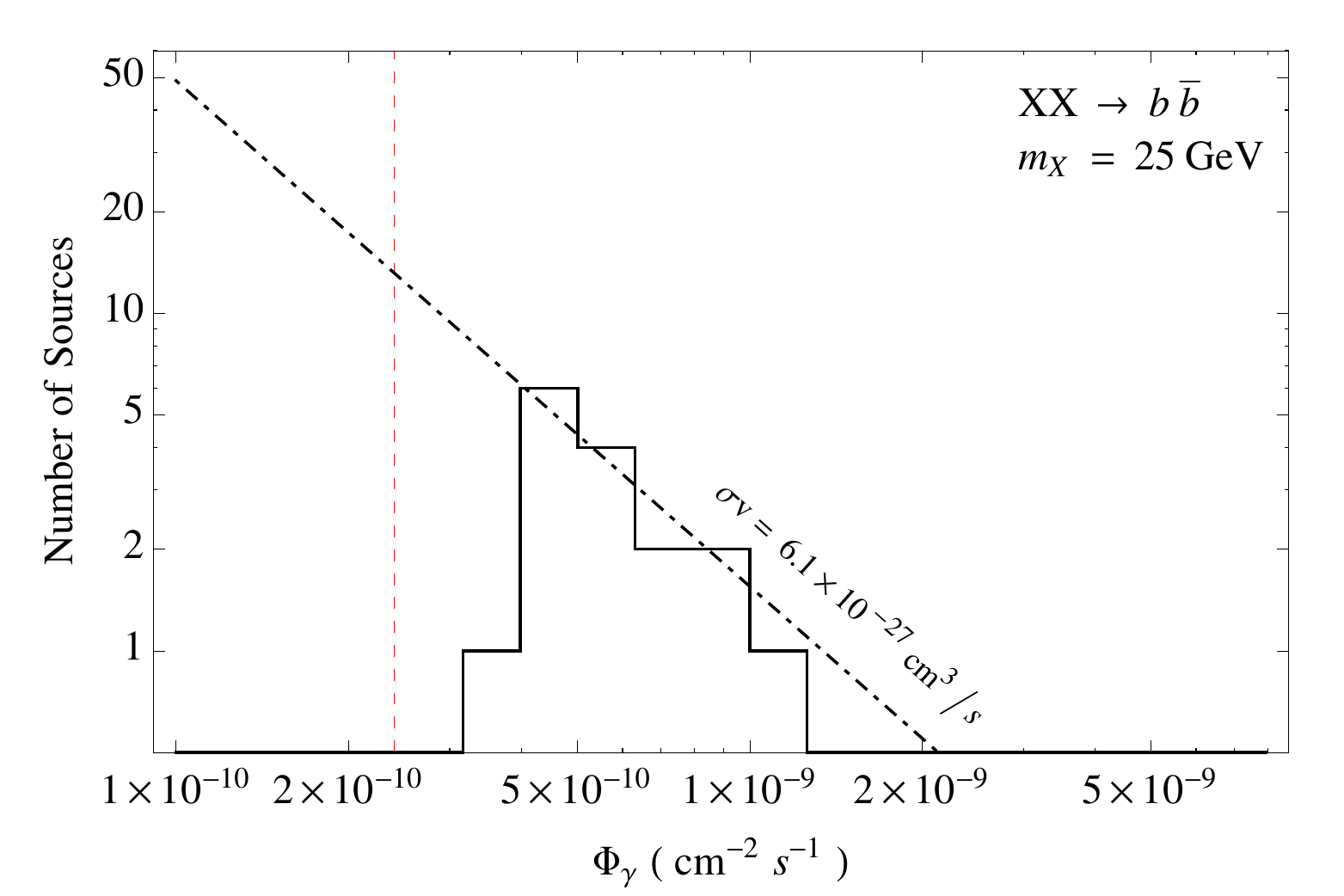}}
\mbox{\includegraphics[width=0.49\textwidth,clip]{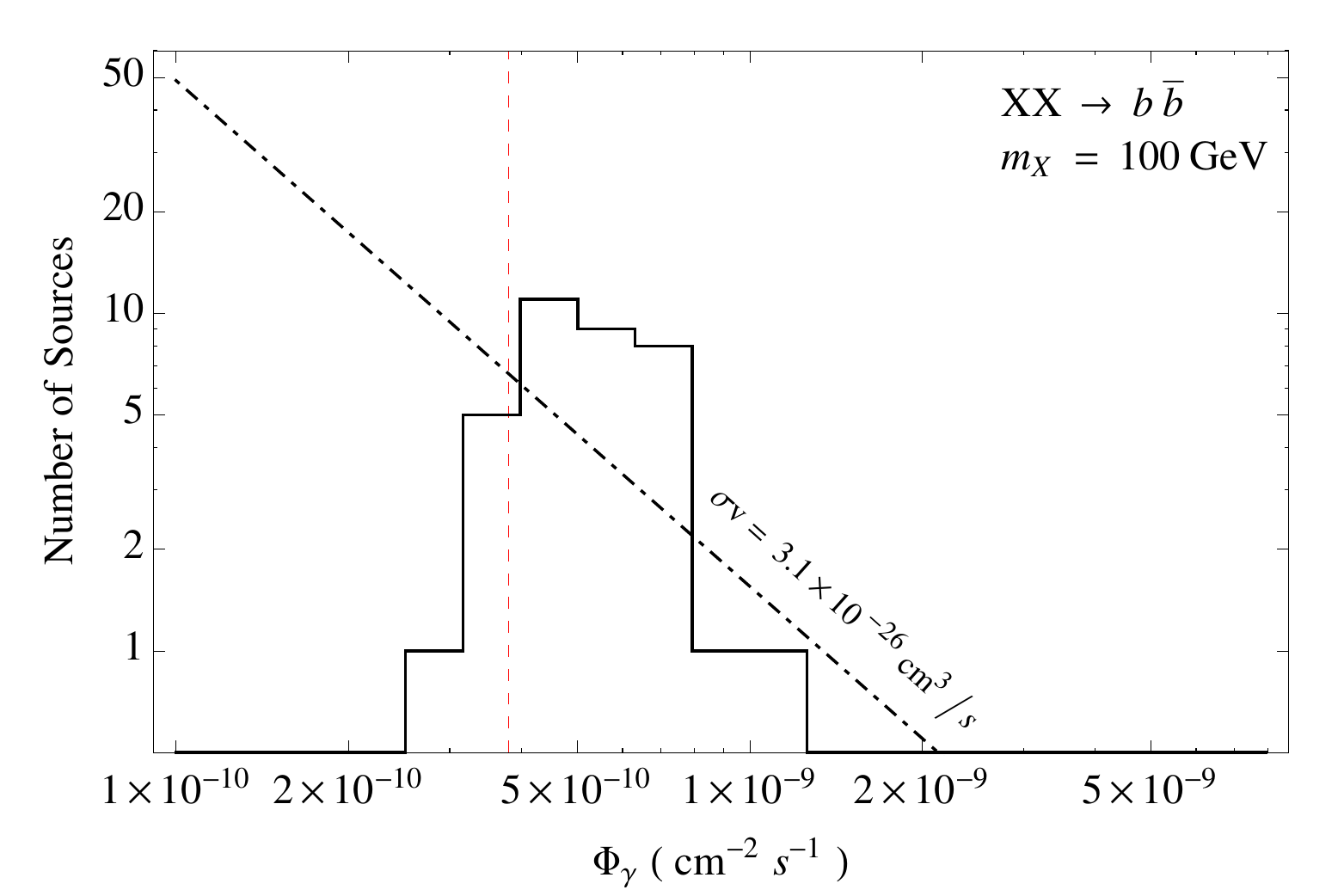}}
\mbox{\includegraphics[width=0.49\textwidth,clip]{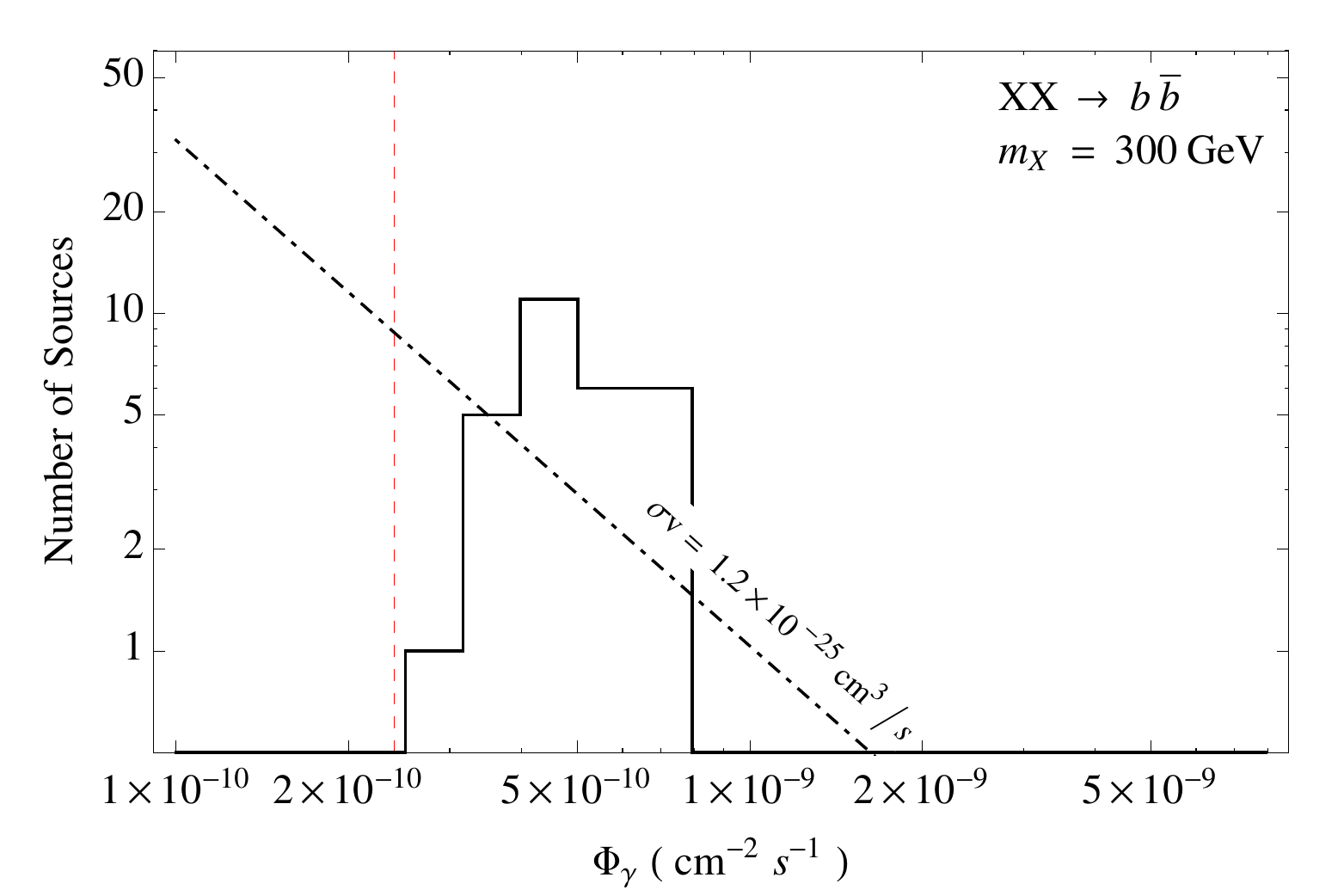}}
\mbox{\includegraphics[width=0.49\textwidth,clip]{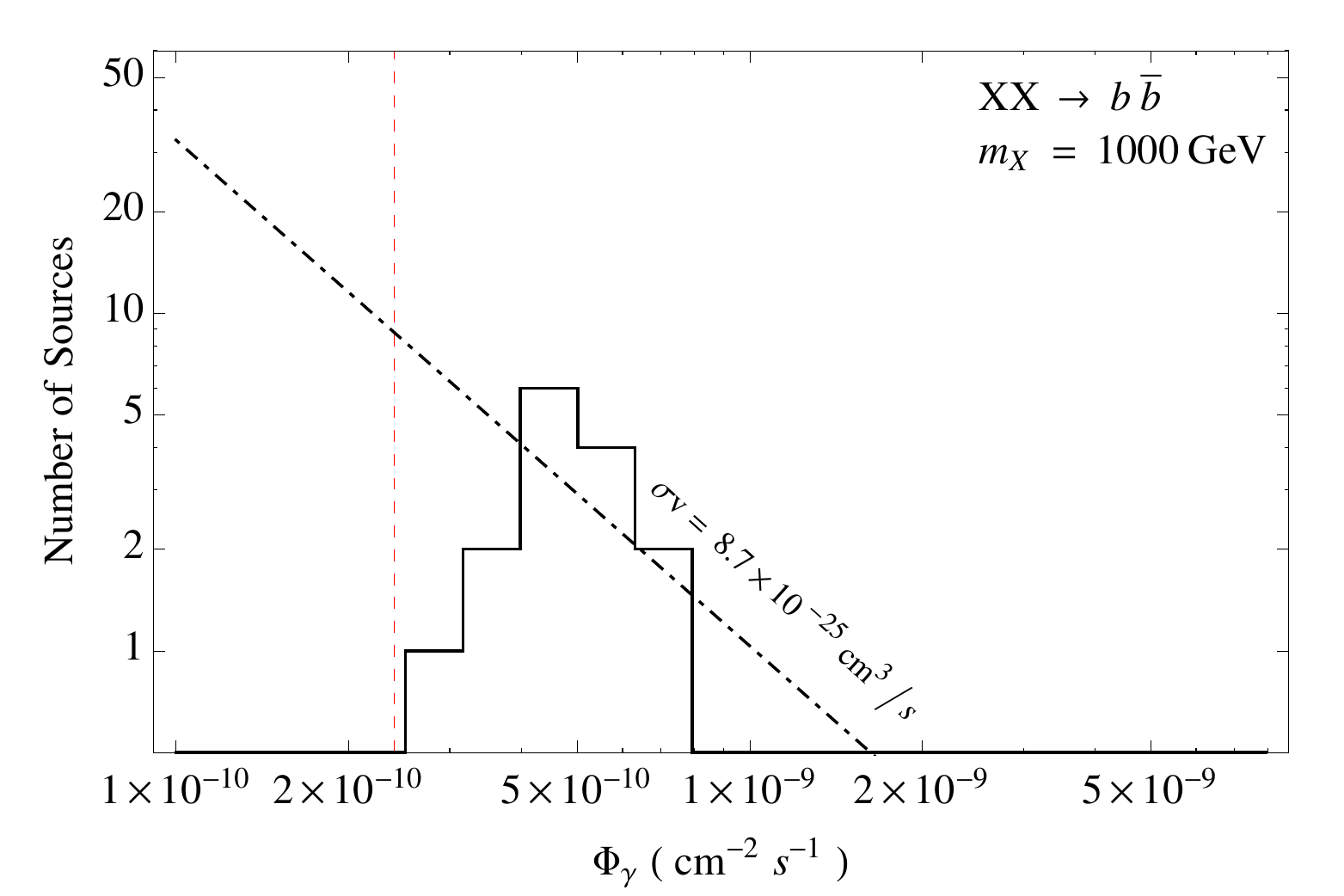}}
\caption{The flux distribution of Fermi's unidentified, non-variable, high latitude ($|b|>30^{\circ}$) sources which are well-fit by dark matter annihilating to $b\bar{b}$ with a masses of 10, 25, 100,  300, or 1000 GeV. In each frame, the vertical red dashed line represents Fermi's approximate point source threshold, for a spectral index chosen to reflect the dark matter mass and annihilation channel under consideration. The dot-dashed curves denote the flux distribution predicted from dark matter subhalos (proportional to $\mathrm{d}N/\mathrm{d}\log_{10}{\Phi_{\gamma}}$) for the value of the annihilation cross section shown. The limits we derive in this study are based on the number of unidentified sources with gamma-ray fluxes greater than $10^{-9}$ photons cm$^{-2}$ s$^{-1}$ (between 1-100 GeV).}
\label{histogramsbb}
\end{figure*}

\begin{figure*}[!]
\mbox{\includegraphics[width=0.49\textwidth,clip]{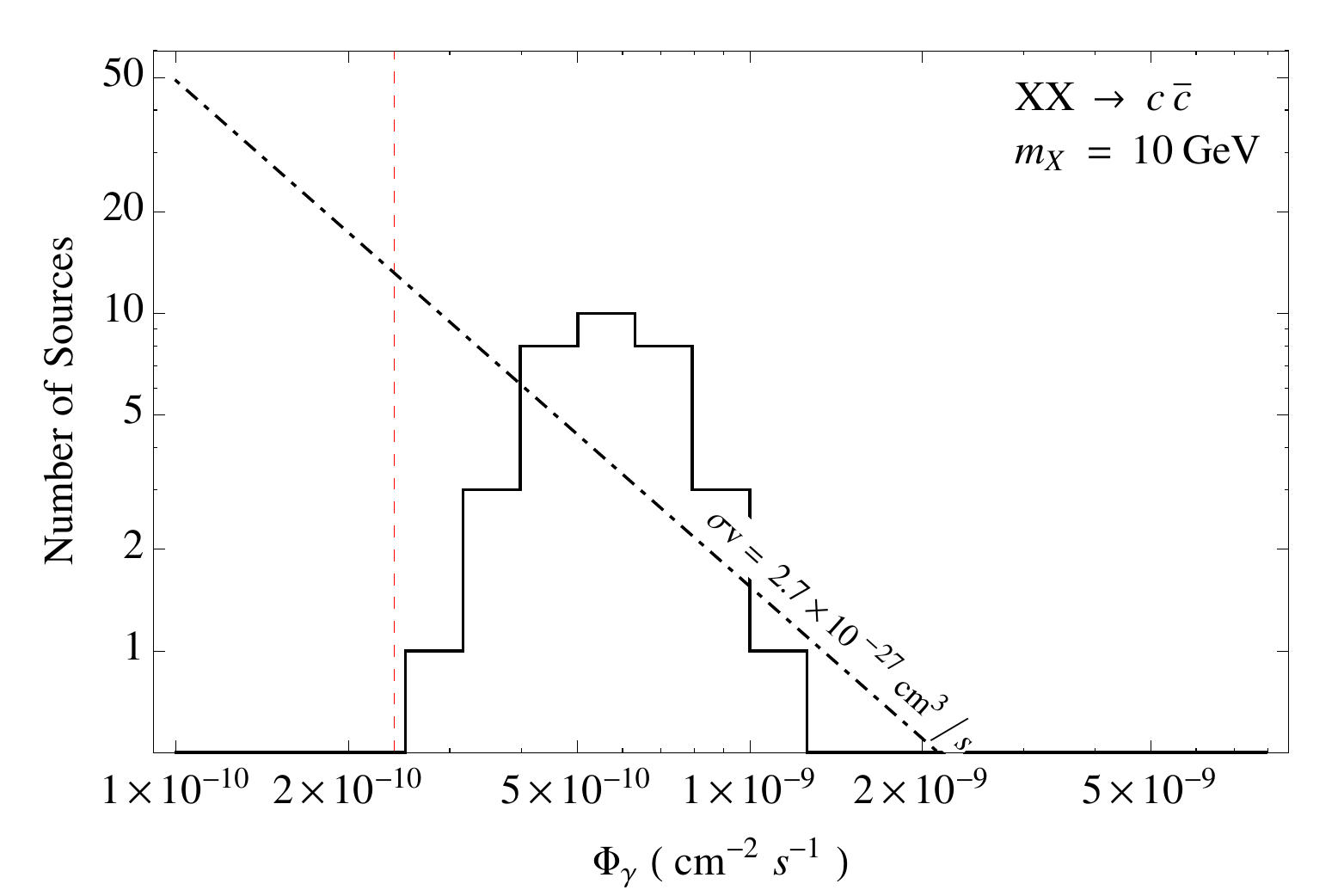}}
\mbox{\includegraphics[width=0.49\textwidth,clip]{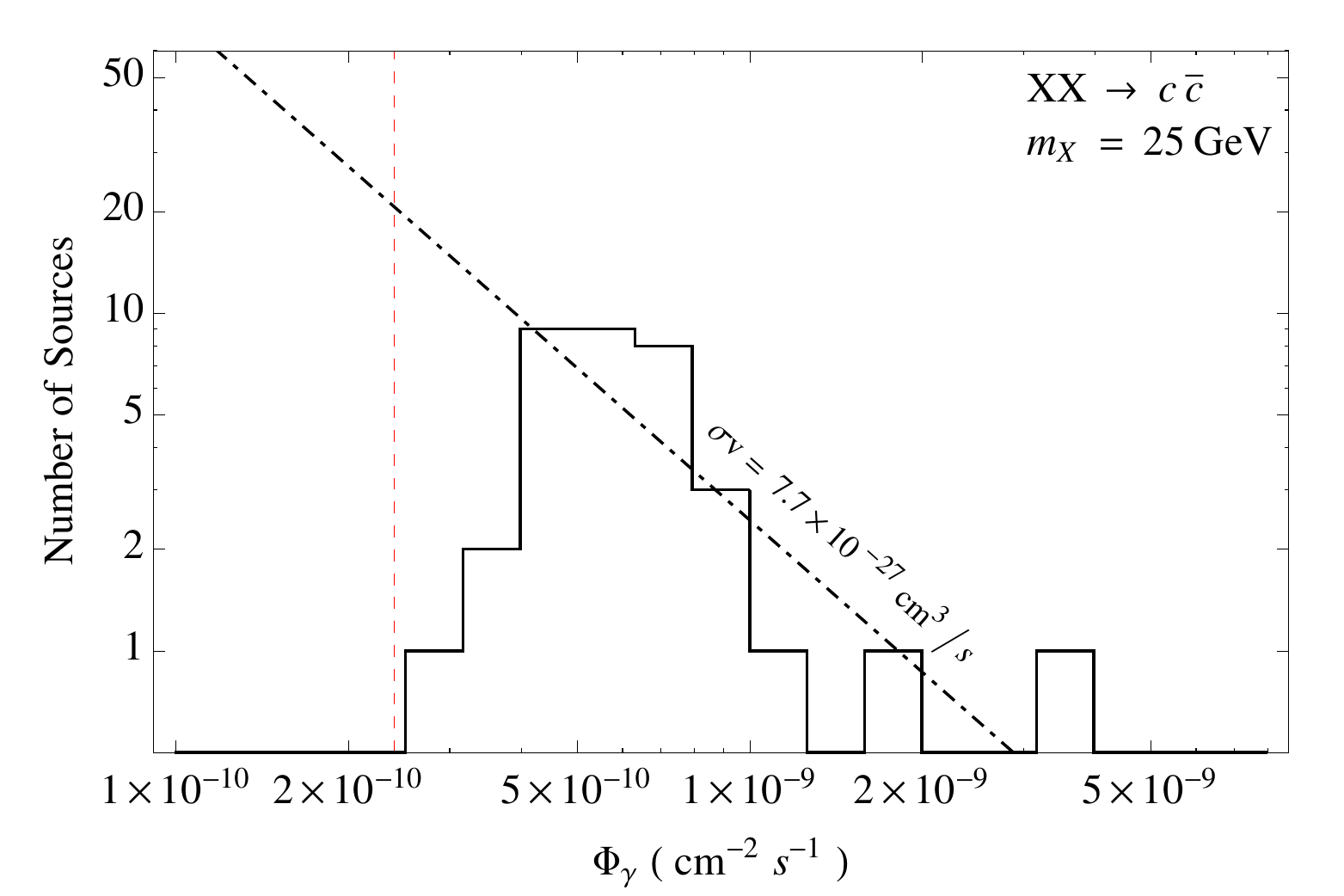}}
\mbox{\includegraphics[width=0.49\textwidth,clip]{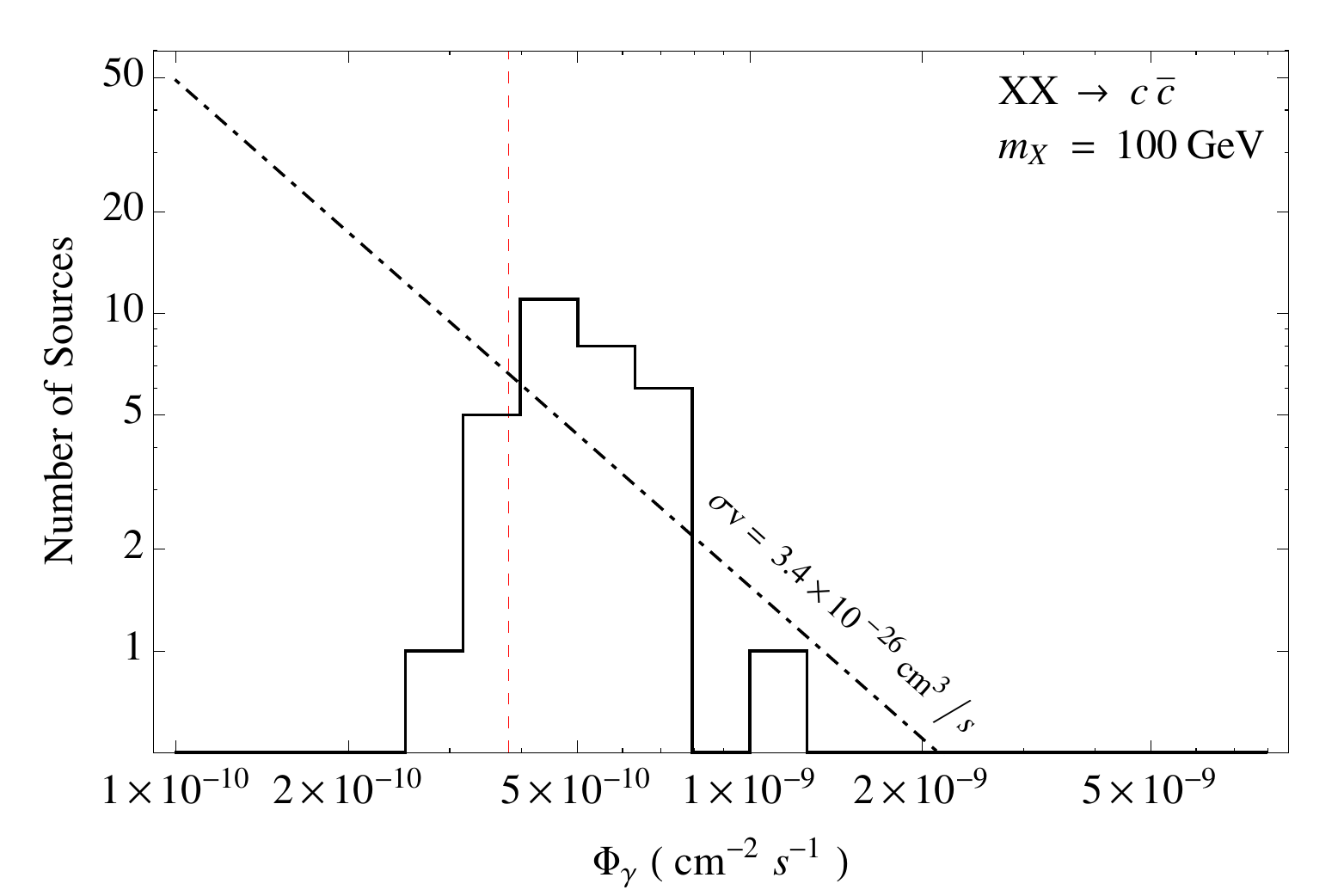}}
\mbox{\includegraphics[width=0.49\textwidth,clip]{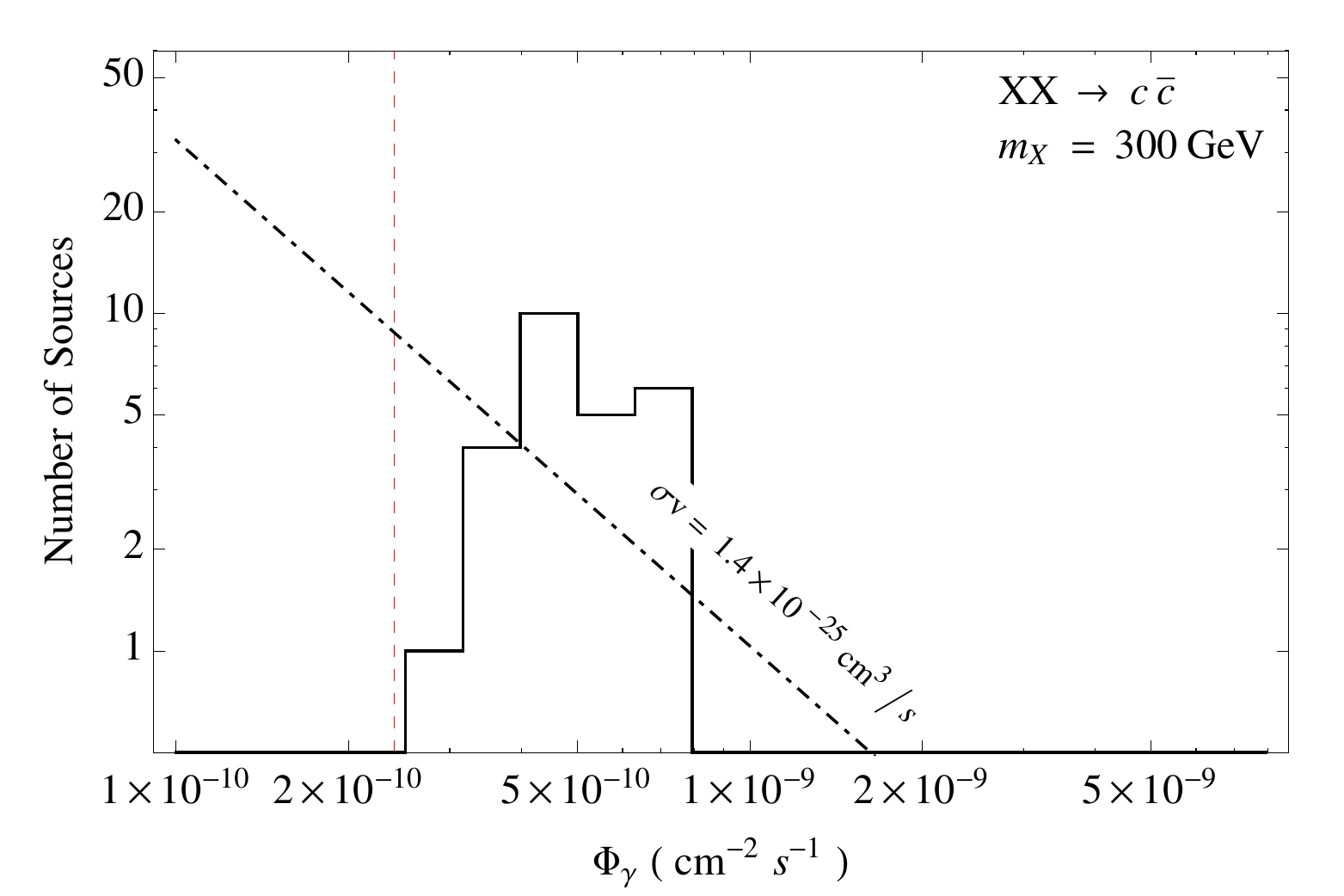}}
\mbox{\includegraphics[width=0.49\textwidth,clip]{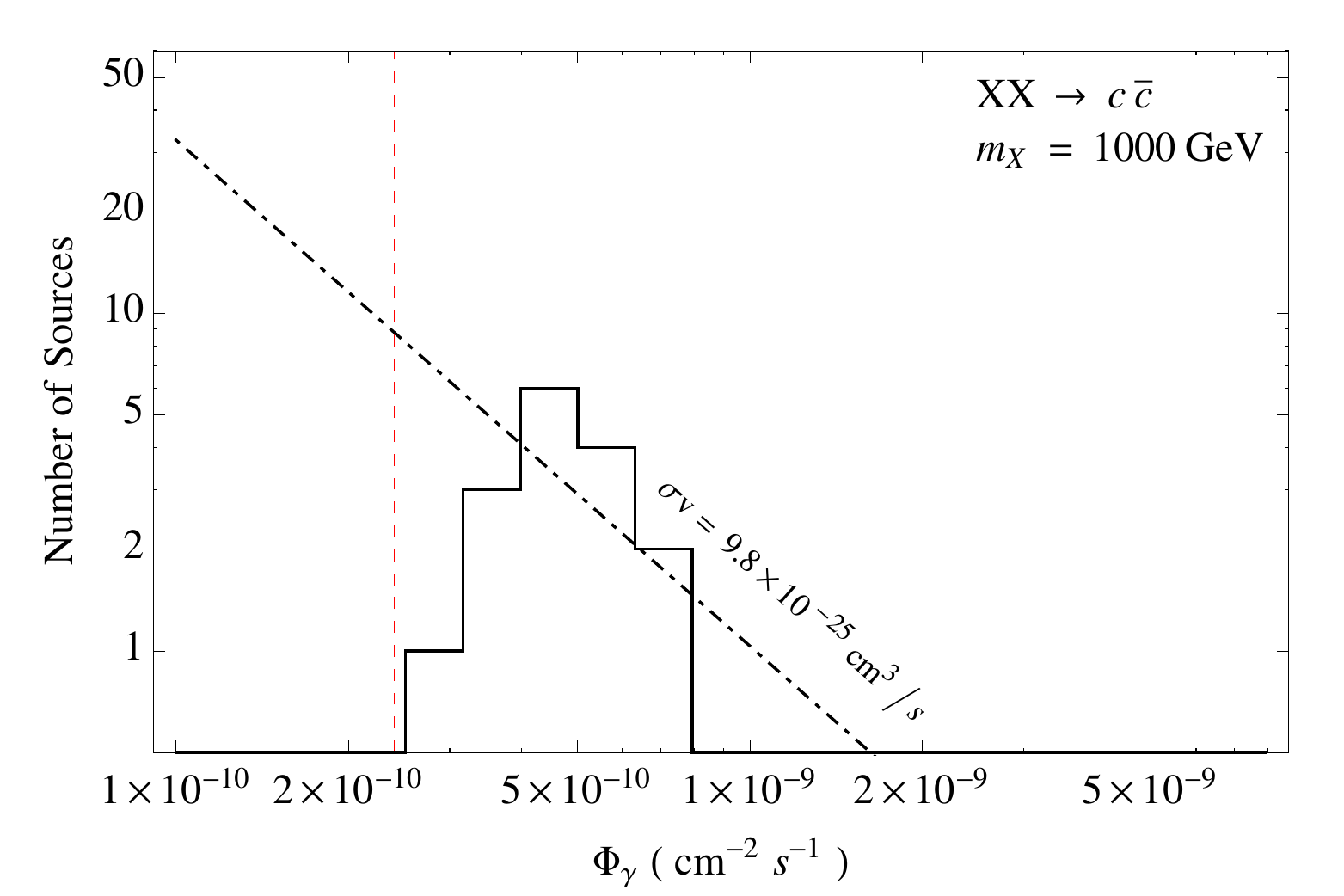}}
\caption{As in Fig.~\ref{histogramsbb}, but for dark matter annihilating to $c\bar{c}$.}
\label{histogramscc}
\end{figure*}

\begin{figure*}[!]
\mbox{\includegraphics[width=0.49\textwidth,clip]{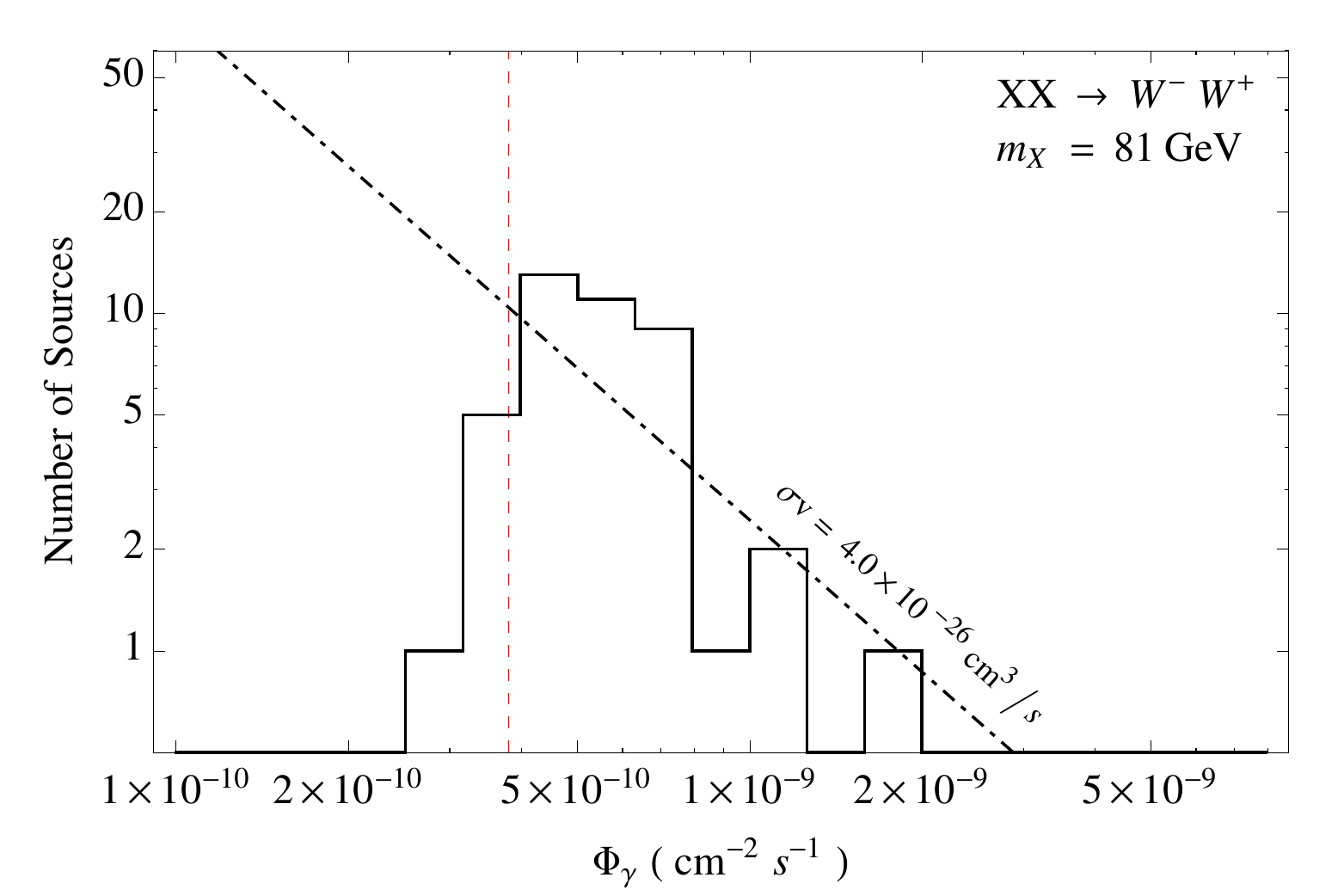}}
\mbox{\includegraphics[width=0.49\textwidth,clip]{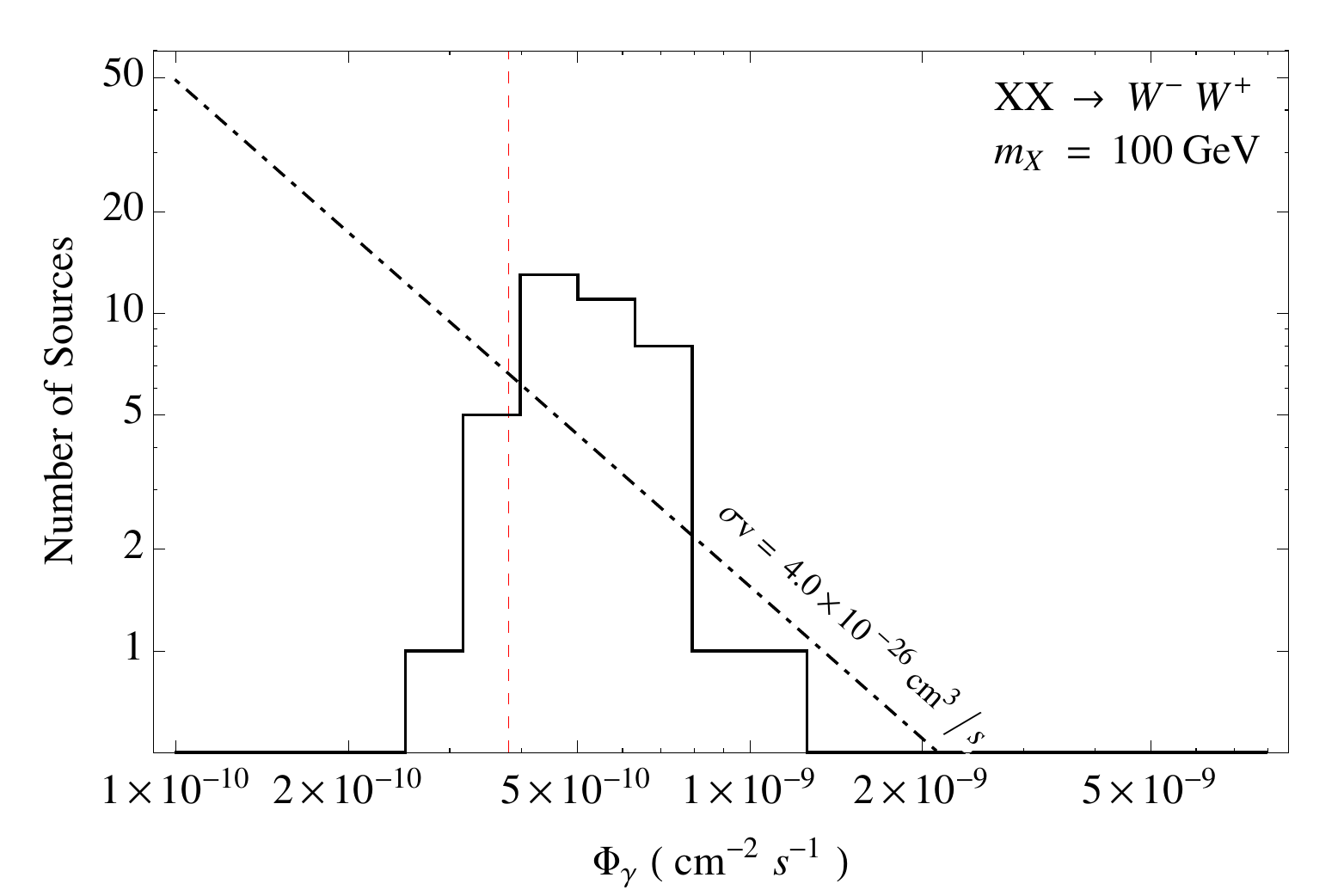}}
\mbox{\includegraphics[width=0.49\textwidth,clip]{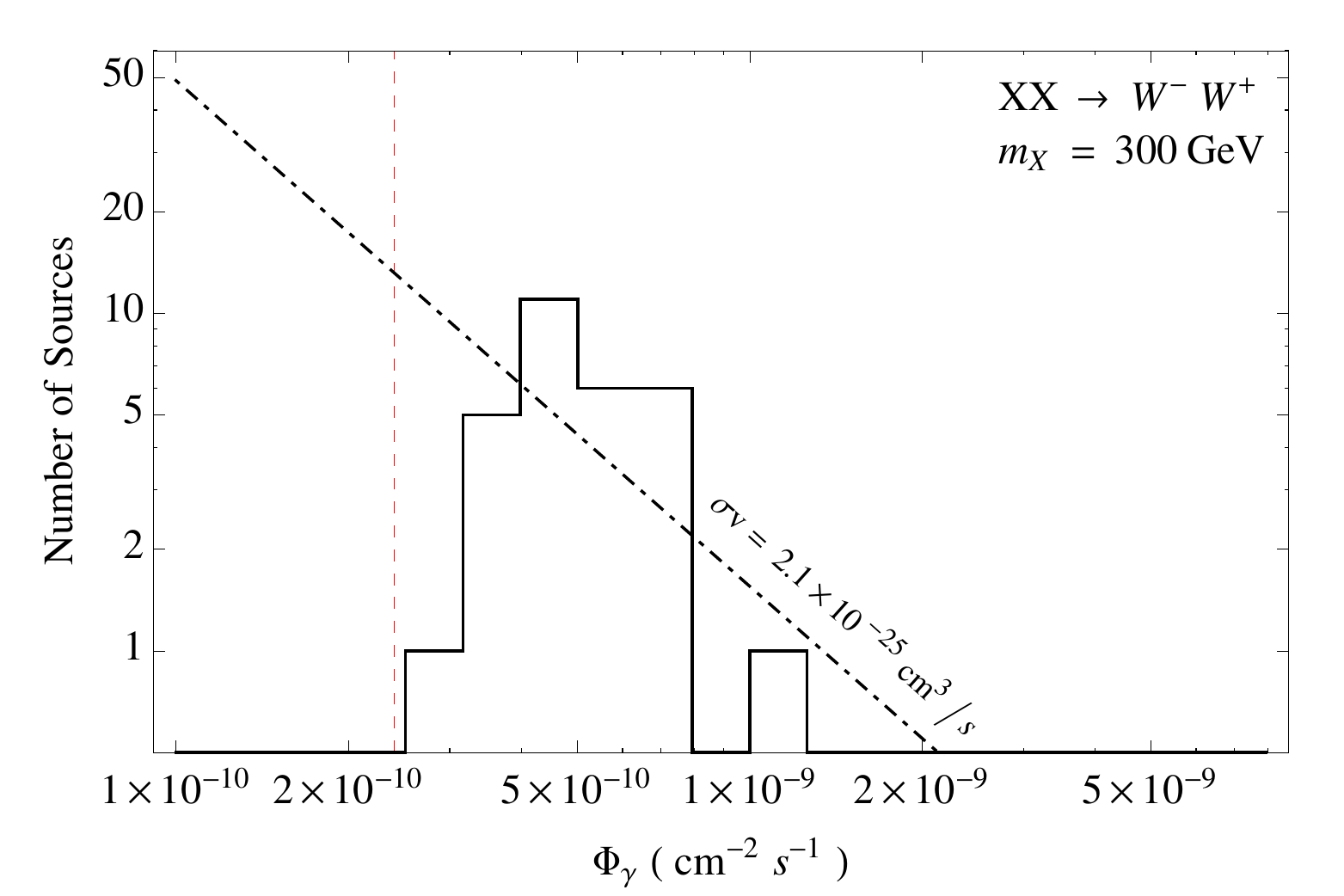}}
\mbox{\includegraphics[width=0.49\textwidth,clip]{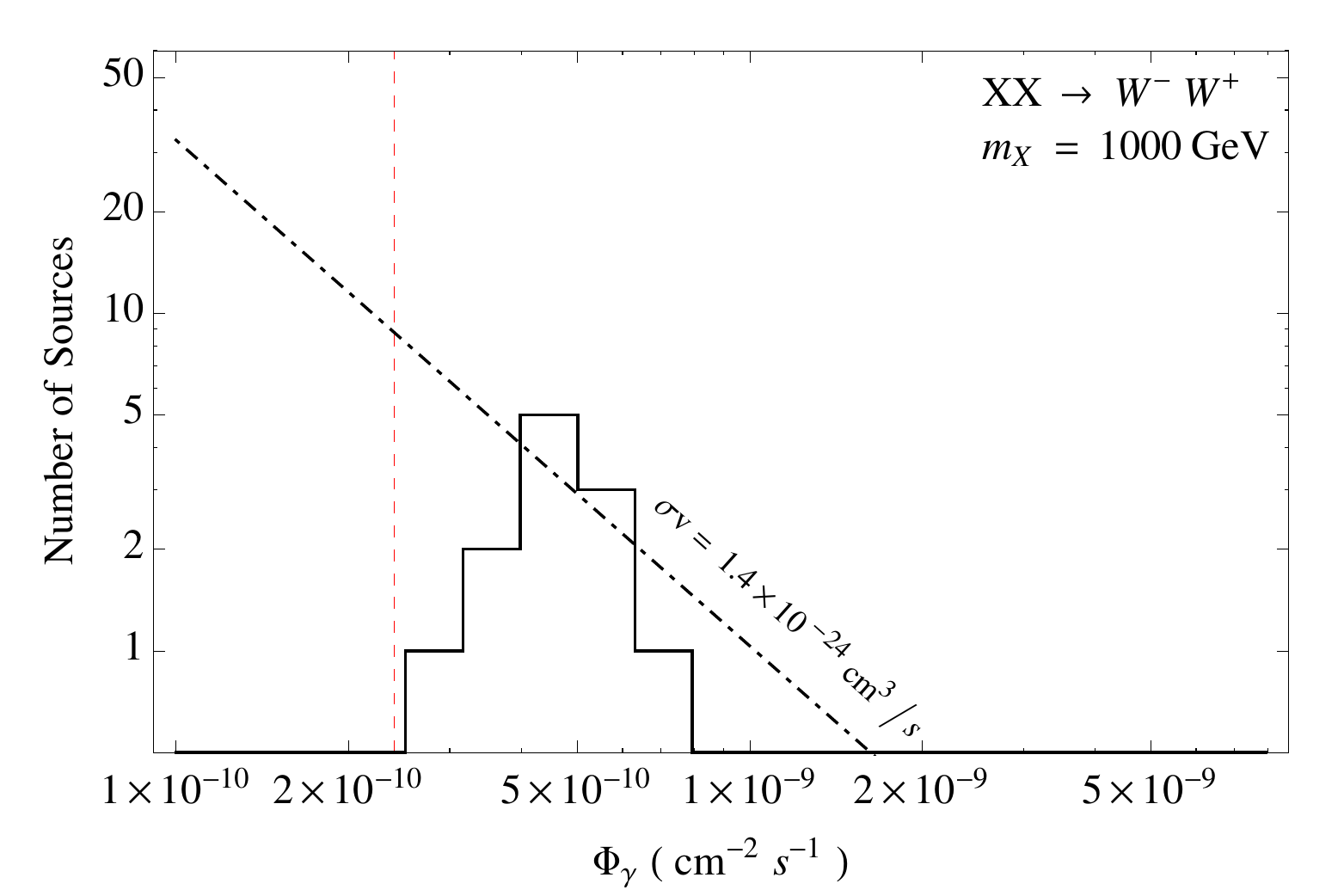}}
\mbox{\includegraphics[width=0.49\textwidth,clip]{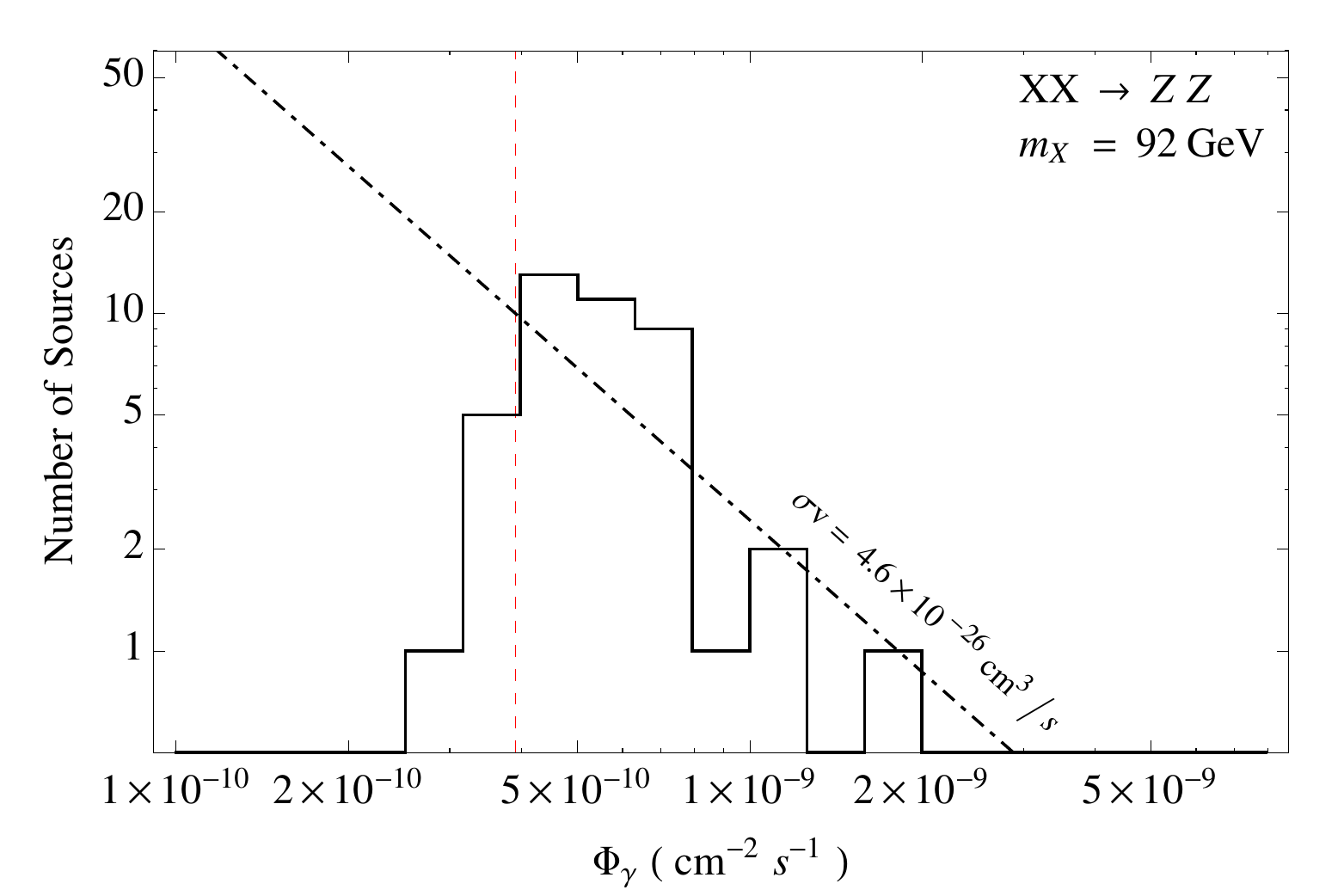}}
\mbox{\includegraphics[width=0.49\textwidth,clip]{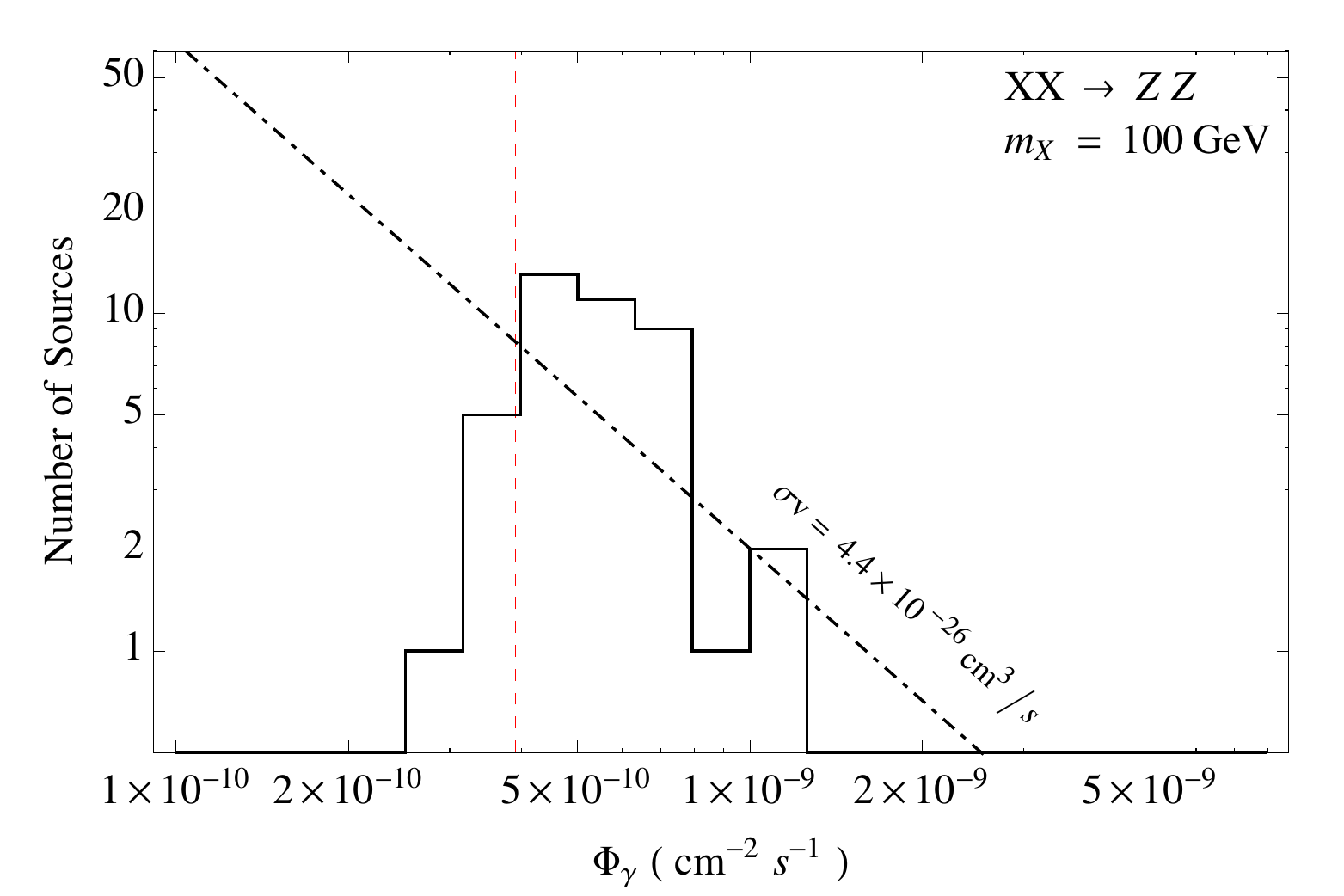}}
\mbox{\includegraphics[width=0.49\textwidth,clip]{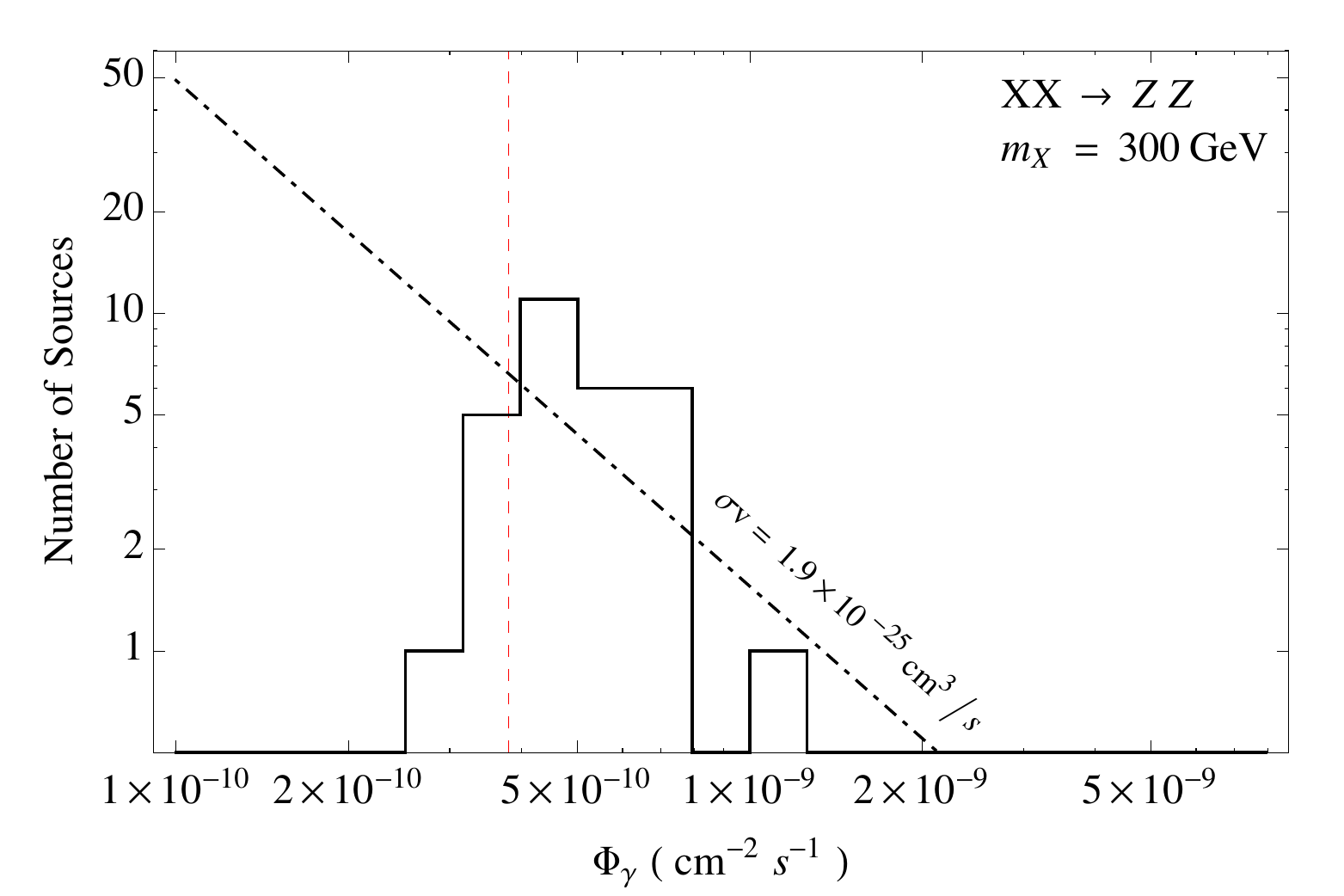}}
\mbox{\includegraphics[width=0.49\textwidth,clip]{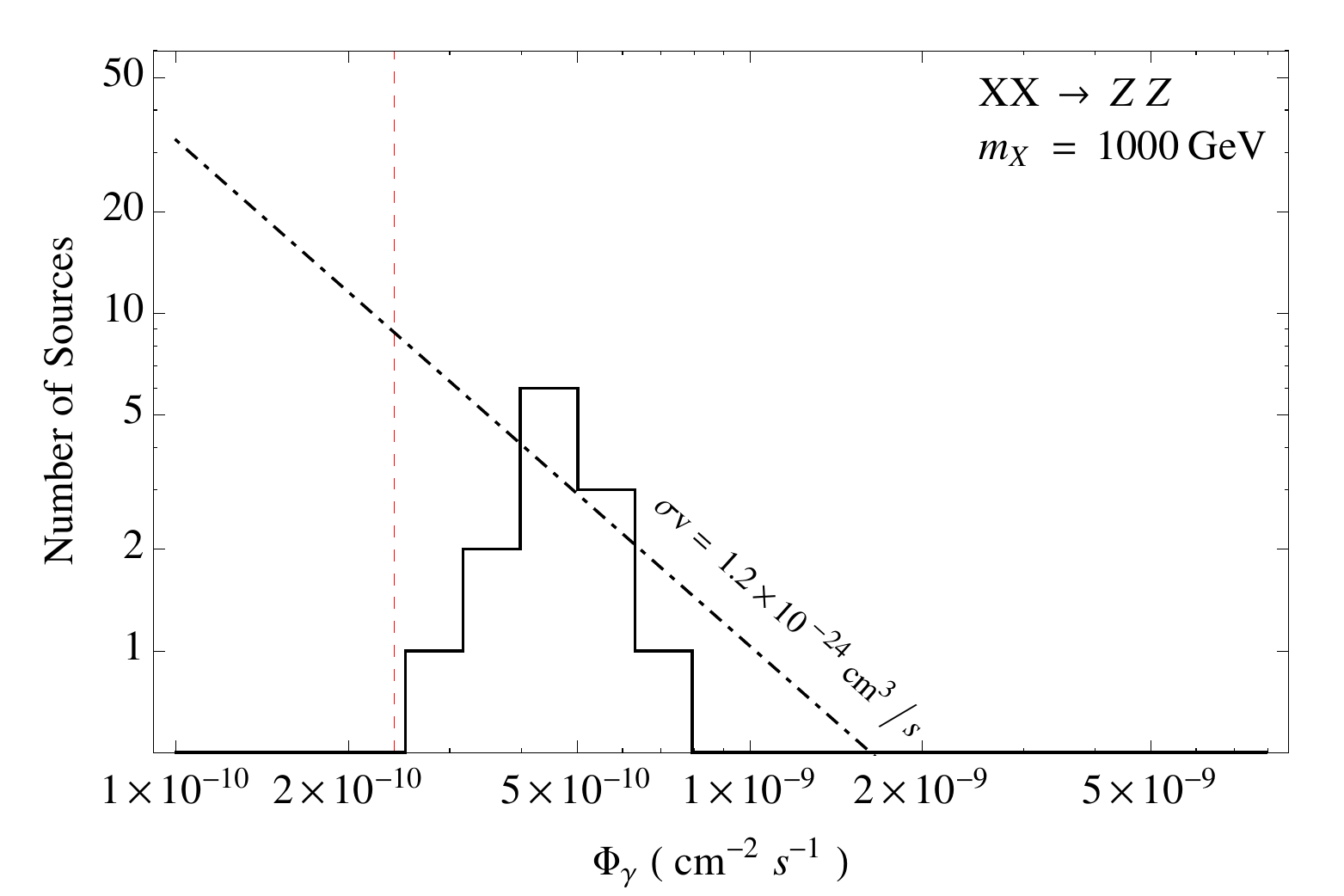}}
\caption{As in Fig.~\ref{histogramsbb}, but for dark matter annihilating to $W^+W^-$ or $ZZ$.}
\label{histogramsww}
\end{figure*}

\begin{figure*}[!]
\mbox{\includegraphics[width=0.49\textwidth,clip]{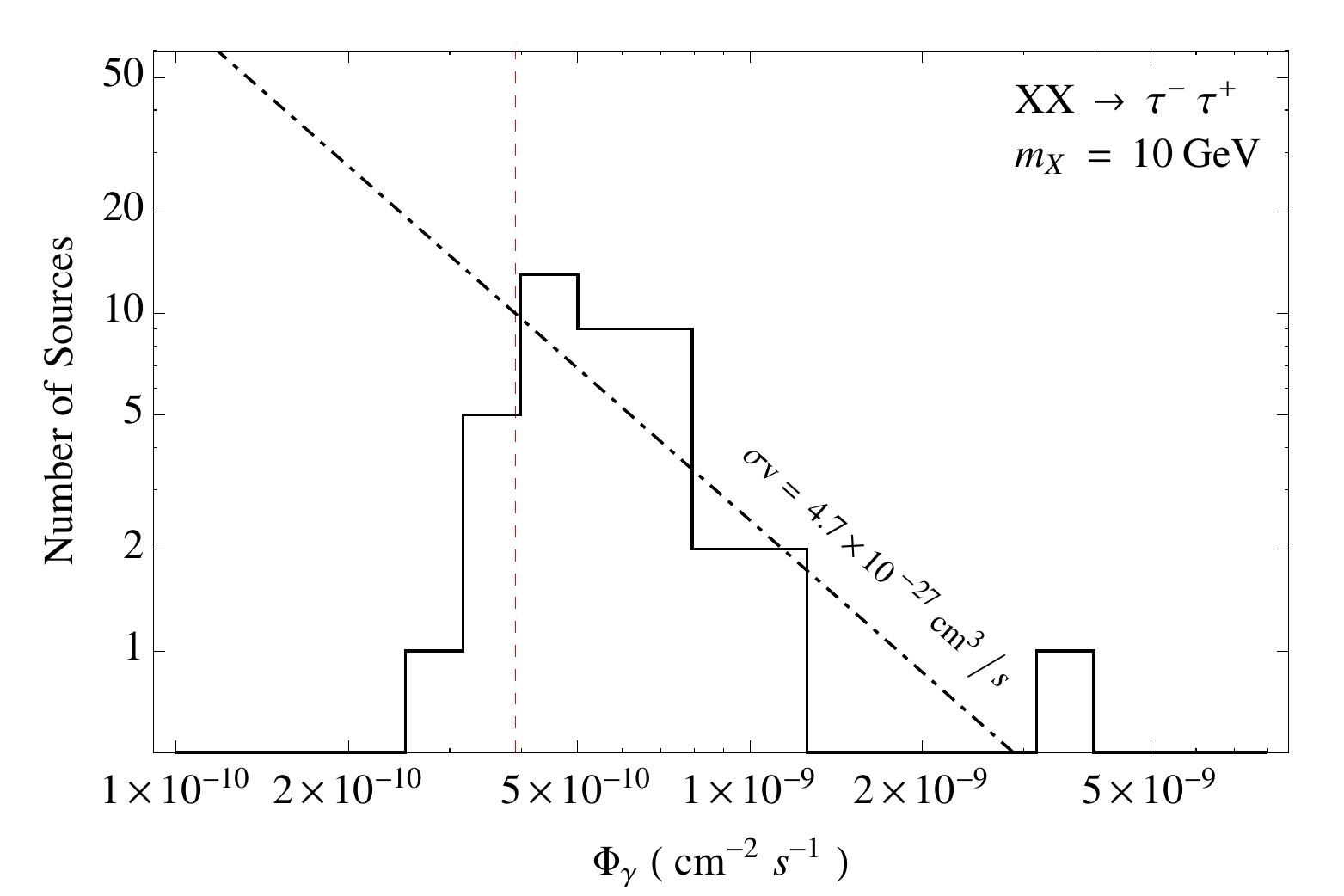}}
\mbox{\includegraphics[width=0.49\textwidth,clip]{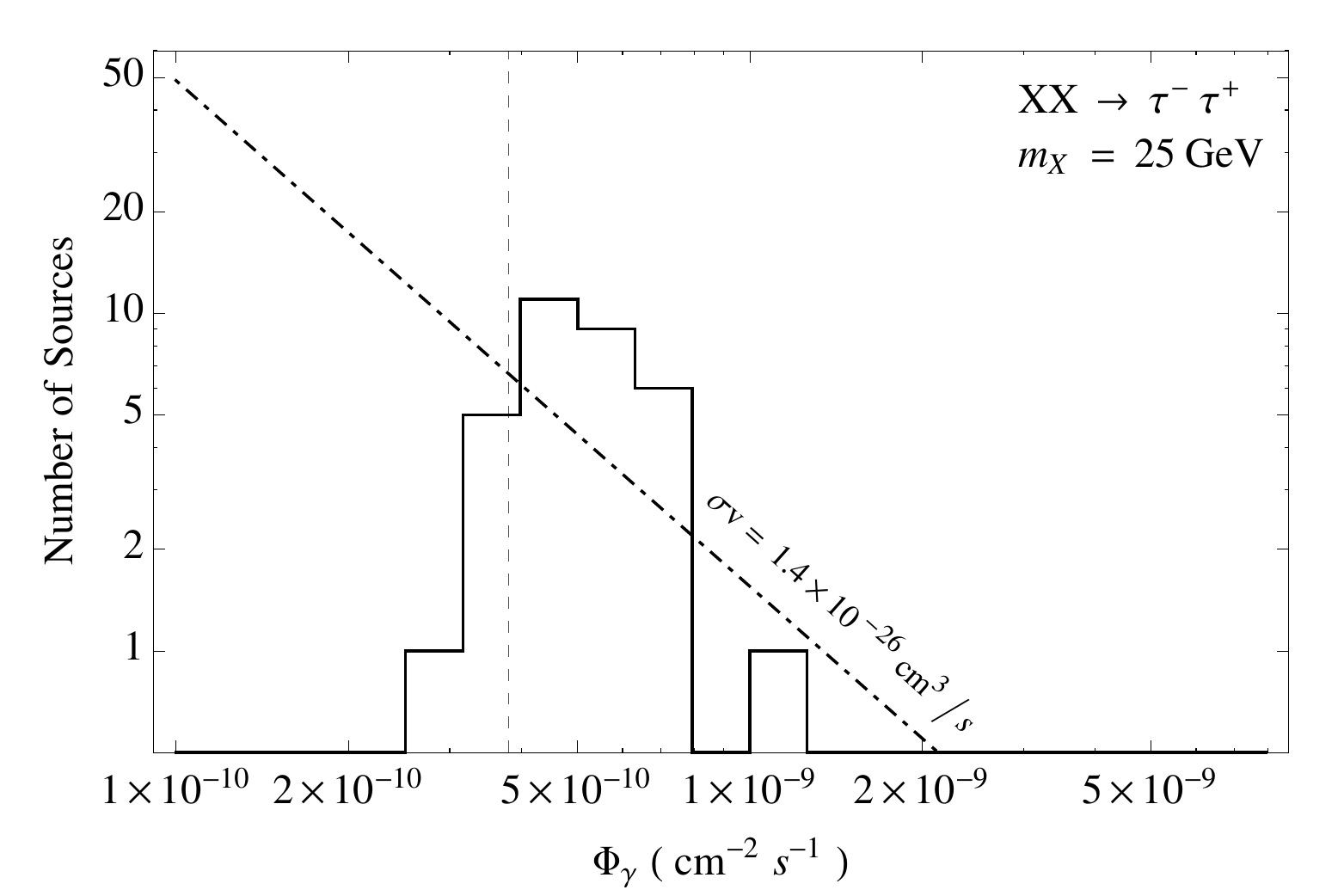}}
\mbox{\includegraphics[width=0.49\textwidth,clip]{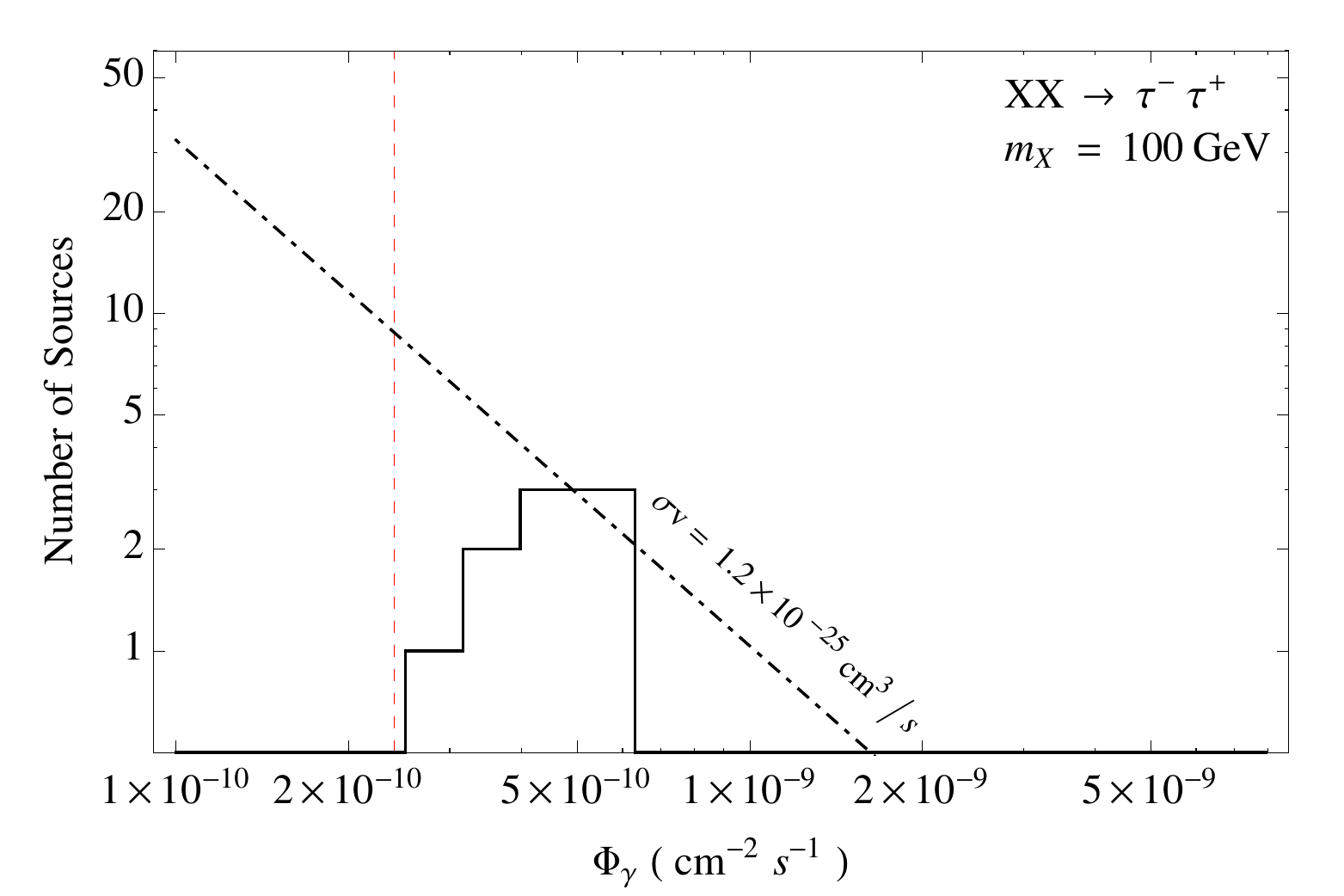}}
\mbox{\includegraphics[width=0.49\textwidth,clip]{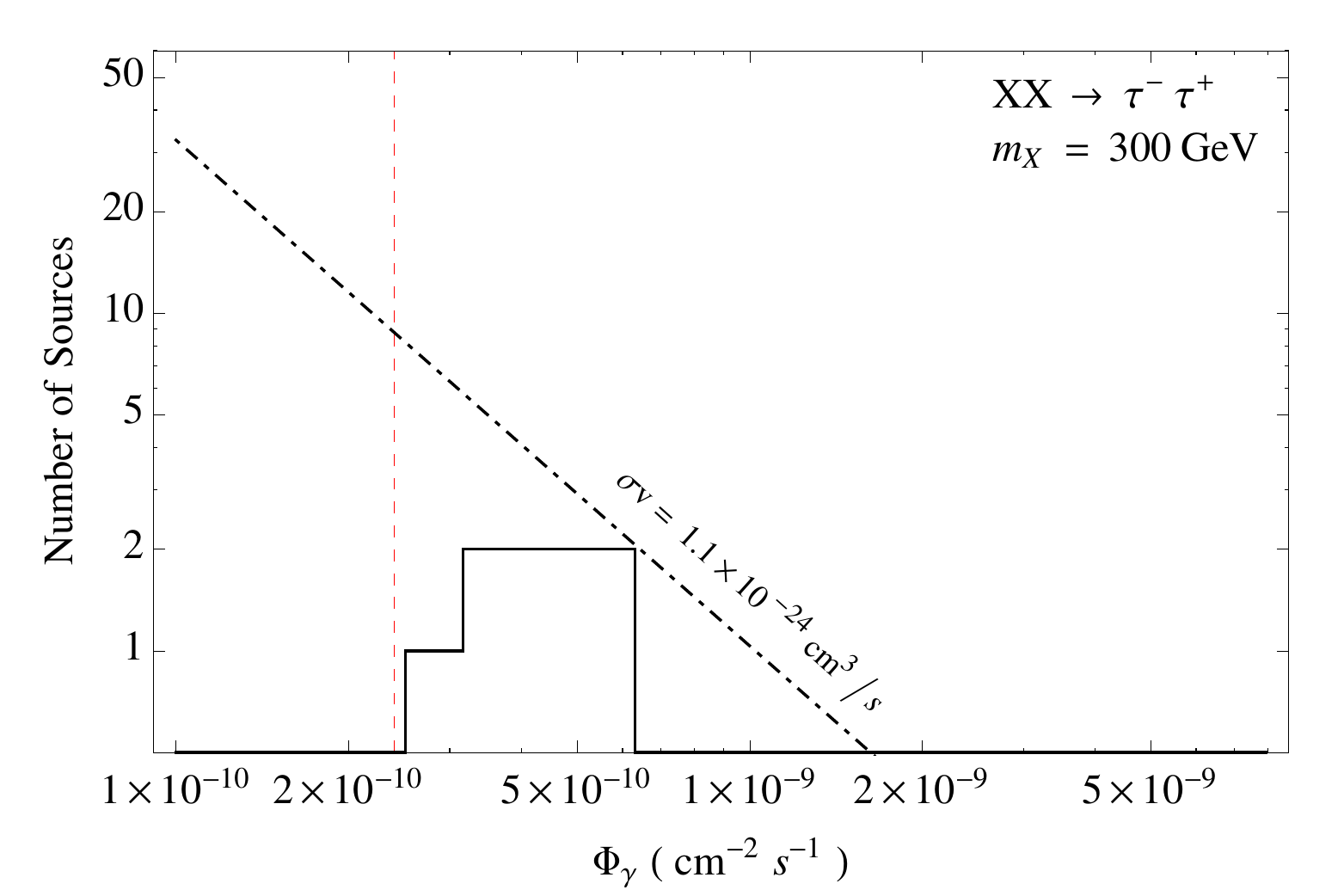}}
\mbox{\includegraphics[width=0.49\textwidth,clip]{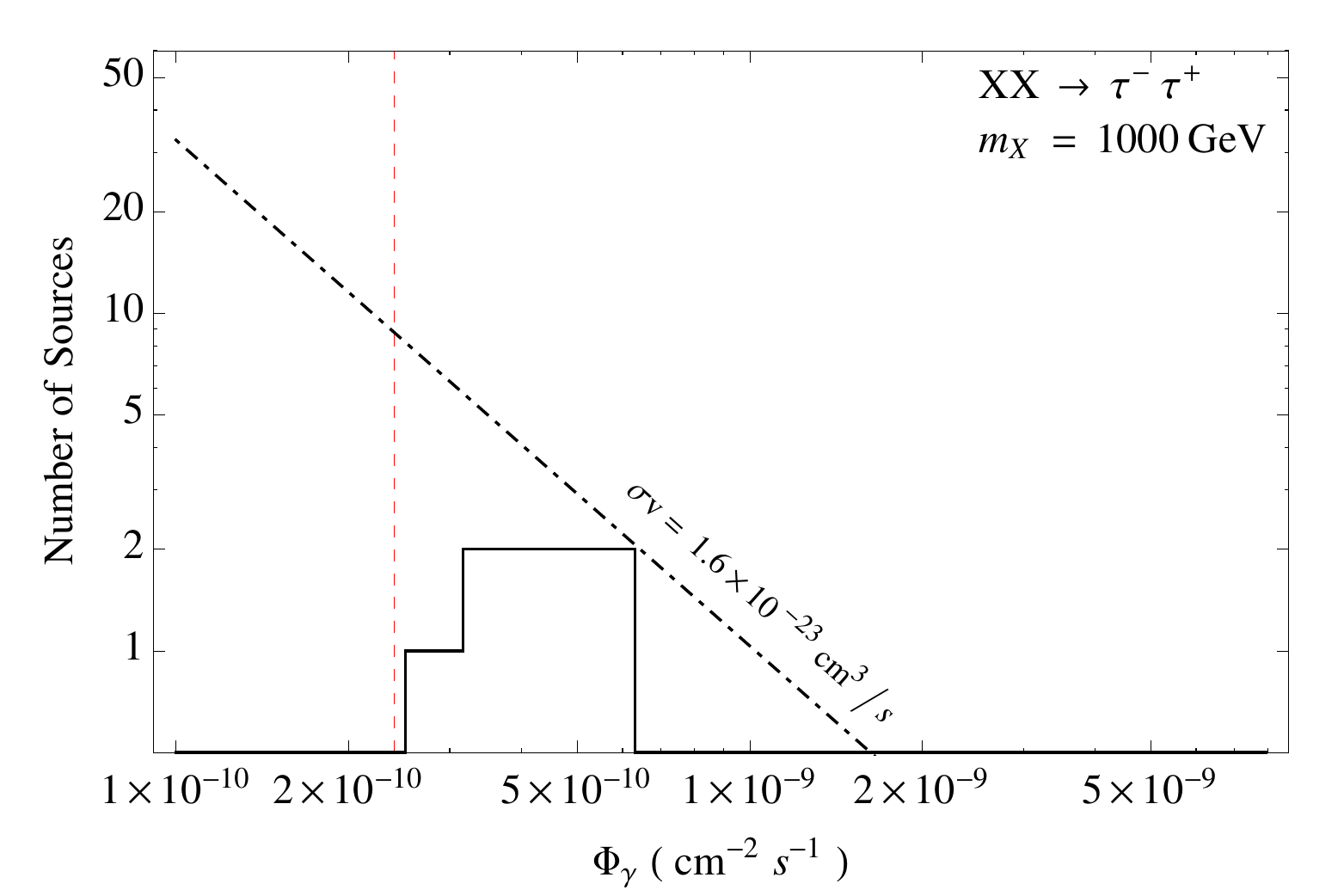}}
\caption{As in Fig.~\ref{histogramsbb}, but for dark matter annihilating to $\tau^+\tau^-$.}
\label{histogramstautau}
\end{figure*}

\begin{figure*}[!]
\mbox{\includegraphics[width=0.49\textwidth,clip]{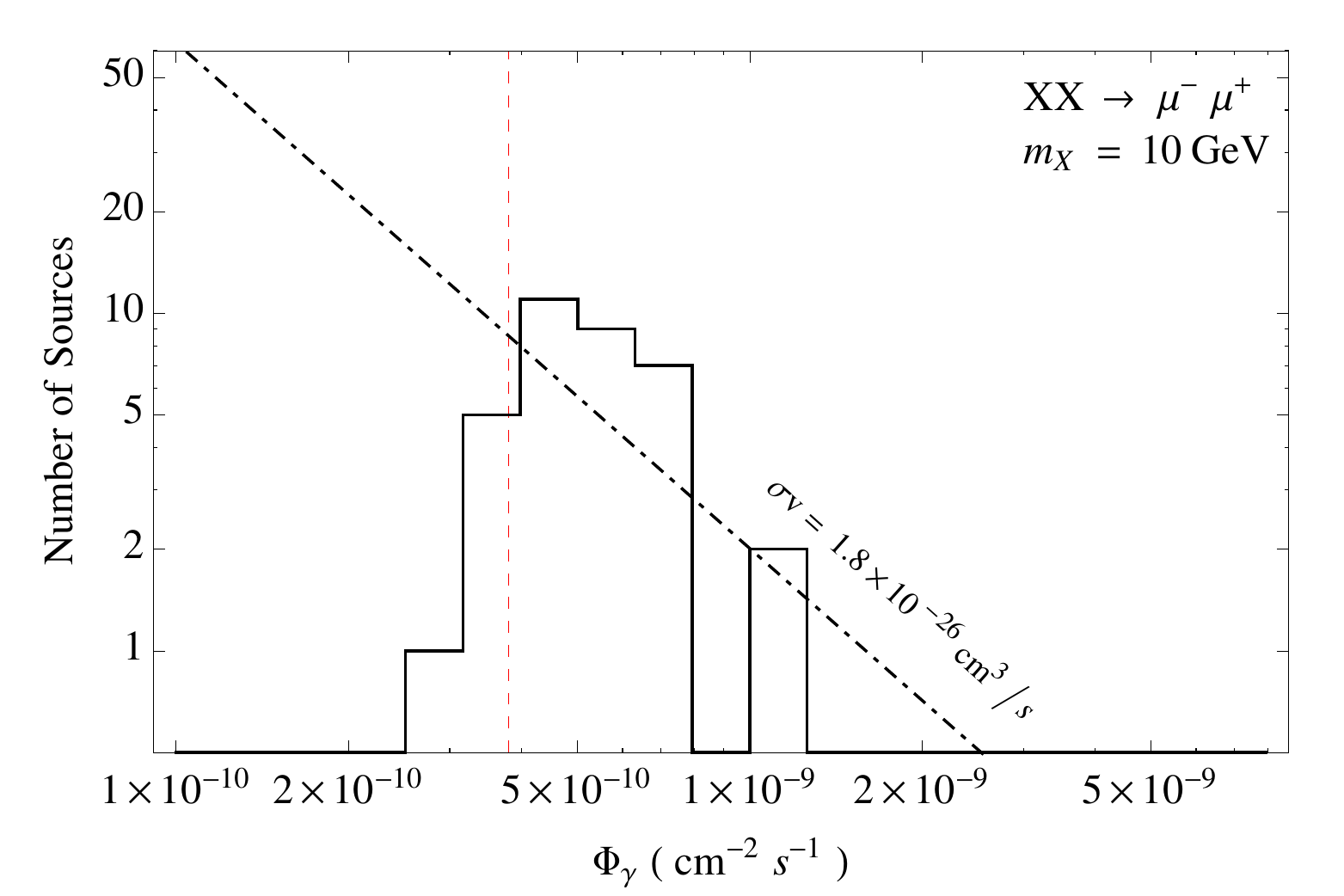}}
\mbox{\includegraphics[width=0.49\textwidth,clip]{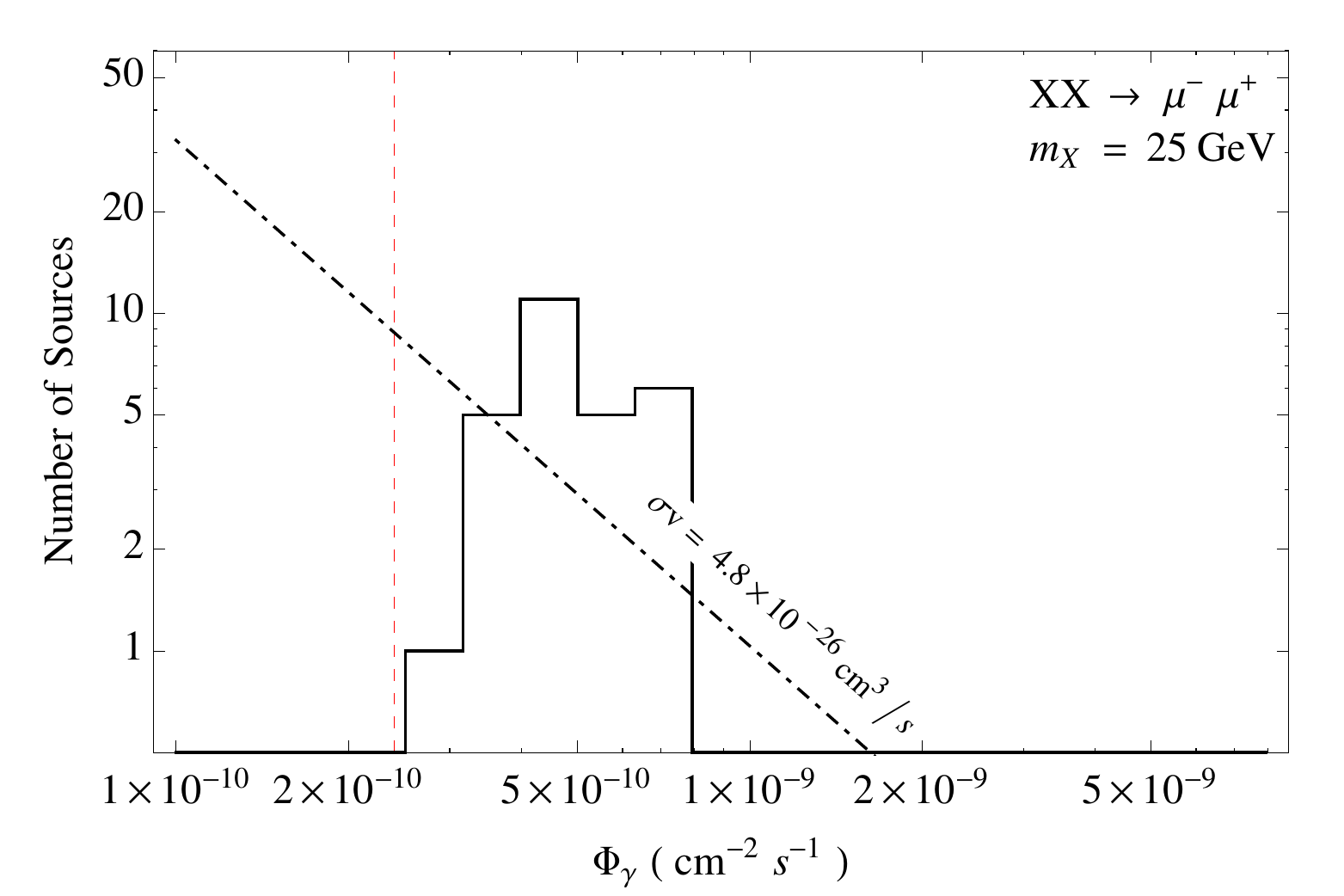}}
\mbox{\includegraphics[width=0.49\textwidth,clip]{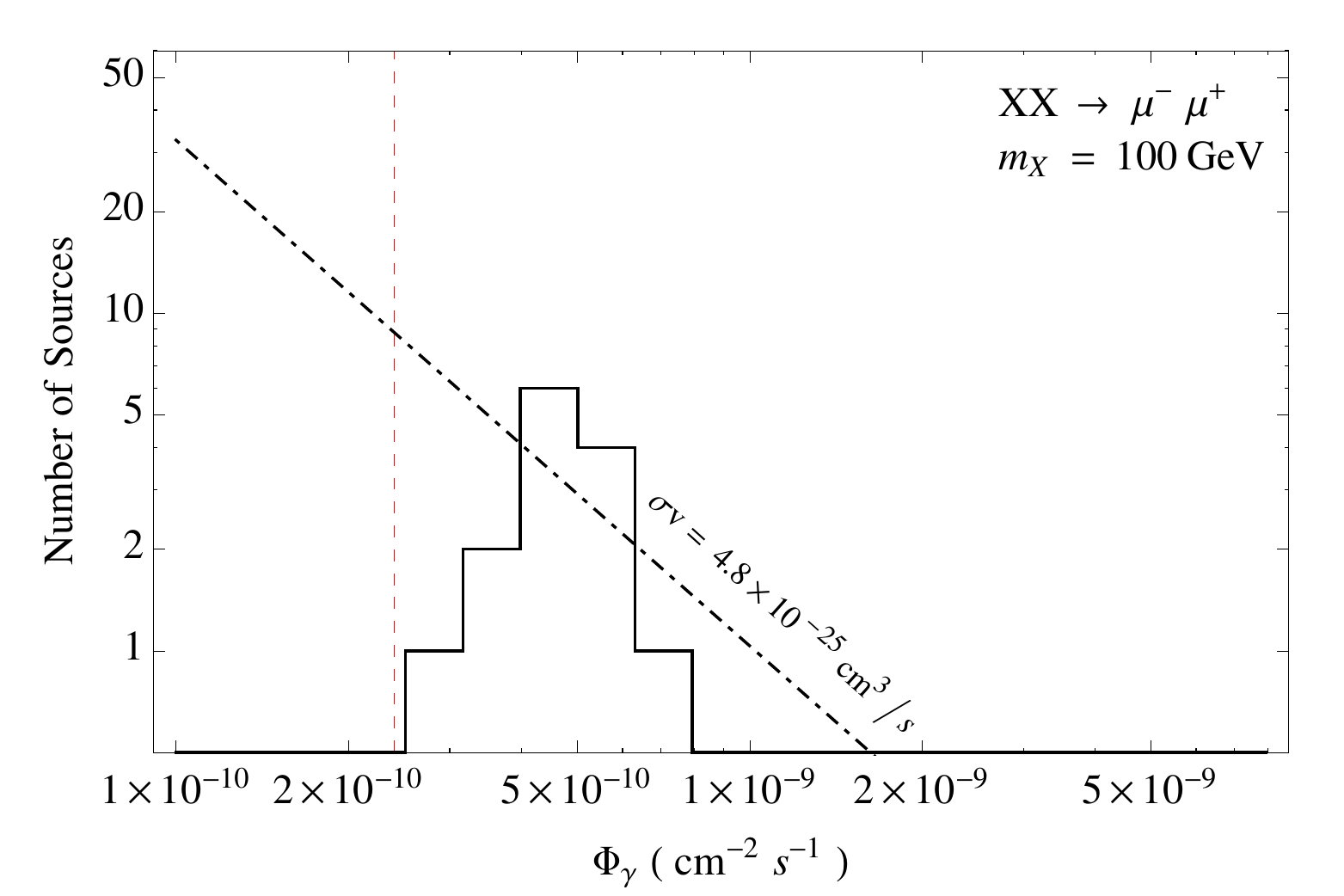}}
\mbox{\includegraphics[width=0.49\textwidth,clip]{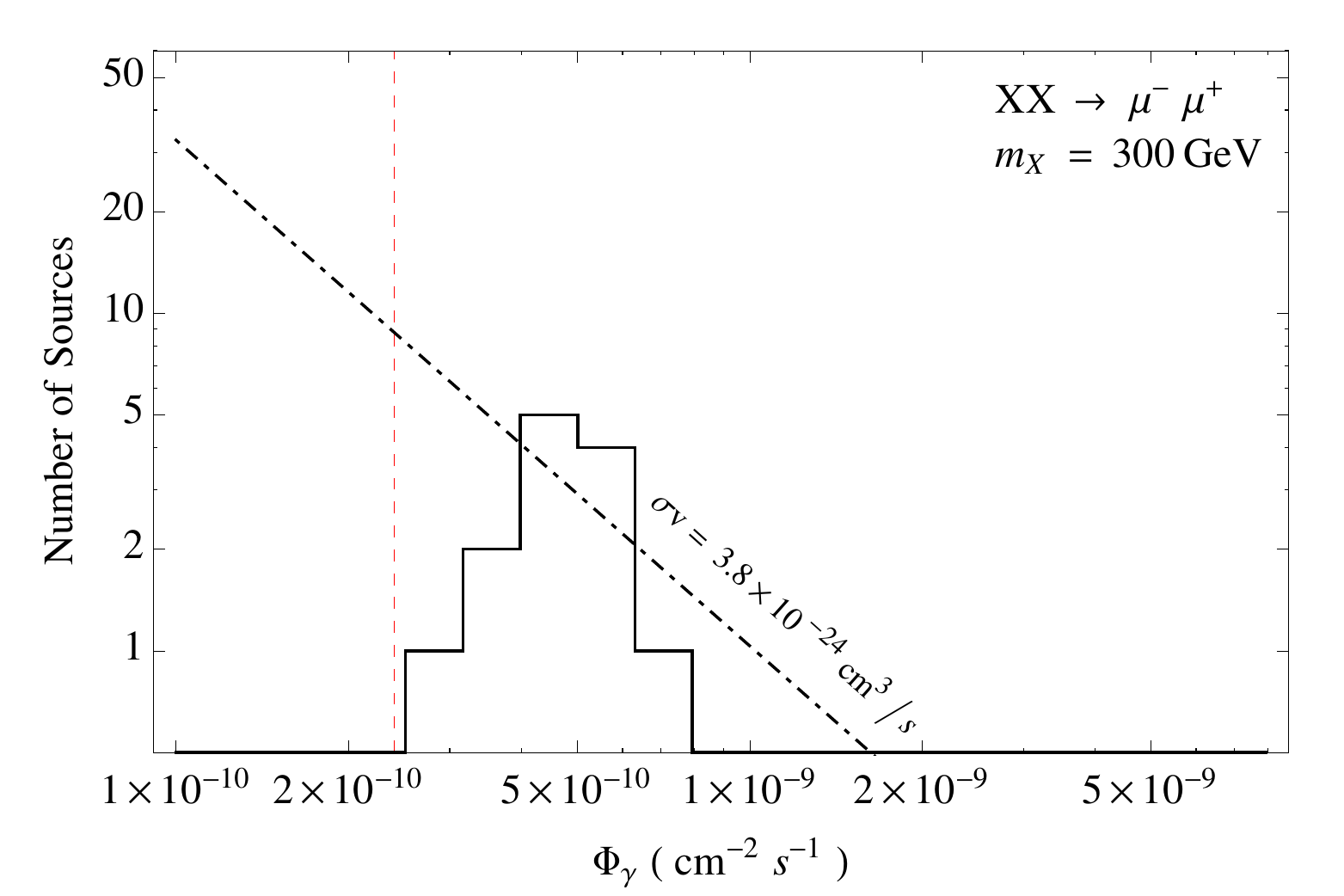}}
\mbox{\includegraphics[width=0.49\textwidth,clip]{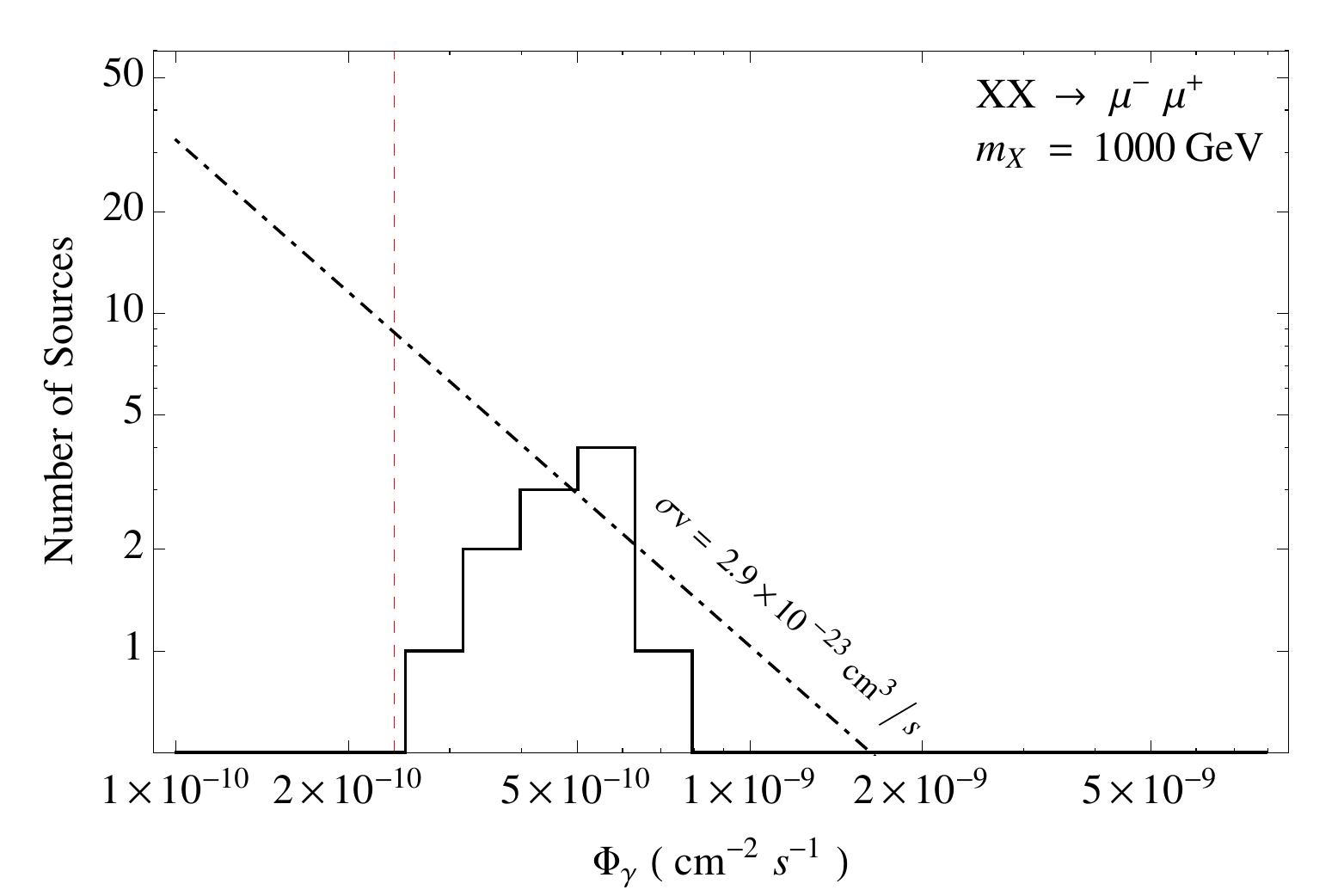}}
\caption{As in Fig.~\ref{histogramsbb}, but for dark matter annihilating to $\mu^+\mu^-$.}
\label{histogramsmumu}
\end{figure*}

In this section, we study the distribution of the unidentified sources observed by Fermi, and use this information to place constraints on the dark matter annihilation cross section. In Figs.~\ref{histogramsbb}-\ref{histogramsmumu}, we plot the flux distribution of Fermi's unidentified, non-variable, high latitude ($|b|>30^{\circ}$) sources that are well-fit by various dark matter models (for several choices of the mass and annihilation channel). We chose to focus on high latitude sources in order to reduce the number of galactic astrophysical sources that contaminate our sample (see, for example, Fig.~2 of Ref.~\cite{Belikov:2011pu}). In each case considered, far fewer sources are shown than in Fig.~\ref{histogramall}, demonstrating that Fermi's unidentified sources exhibit a wide range of spectral shapes.  No single dark matter model can account for more than about half of these sources. 

Also shown as a dot-dashed line in each frame of Figs.~\ref{histogramsbb}-\ref{histogramsmumu} is the flux distribution predicted from dark matter subhalos (proportional to $\mathrm{d}N/\mathrm{d}\log_{10}{\Phi_{\gamma}}$) for the value of the annihilation cross section shown. In each case, this curve exceeds the number of very bright ($>10^{-9}$ cm$^{-2}$ s$^{-1}$ between 1-100 GeV) unidentified sources, but often falls below the number of observed sources with lower fluxes.  For this reason, we find that we can derive the strongest possible limits on the annihilation cross section by focusing on the brightest of Fermi's unidentified sources.  

For each choice of dark matter mass and annihilation channel, we count the number of well-fit sources with $\Phi_{\gamma}>10^{-9}$ cm$^{-2}$ s$^{-1}$ and use this to determine the 95\% confidence level poisson upper limit on the predicted number of such sources. We then determine the annihilation cross section which corresponds to this number of sources. In Fig.~\ref{limits}, we show the resulting upper limits on the dark matter annihilation cross section, as a function of mass, and for several choices of the annihilation channel.

In addition to this result derived from high-latitude, unidentified, point-like gamma-ray sources, we can also use the lack of any observed spatially extended dark matter subhalo candidates to place constraints on the dark matter annihilation cross section. To do so, we have calculated the number of subhalos predicted outside of the Galactic Plane ($|b|>10^{\circ}$) with a flux exceeding Fermi's threshold for detectable spatial extension (see the right frame of Fig.~\ref{detect}), for a given dark matter mass, annihilation channel, and cross section. This is shown in the left frame of Fig.~\ref{extendedlimits} for the case of a 100 GeV particle annihilating to $b\bar{b}$. Given that zero extended subhalo candidates have been observed, we translate this result into a limit by determining the annihilation cross section that predicts 2.996 such sources (the 95\% poisson upper limit for an observation of zero). We show the resulting limit (for annihilations to $b\bar{b}$) in the right frame of Fig.~\ref{extendedlimits}. Comparing this to the results shown in Fig.~\ref{limits}, we see that the constraint derived from the lack of observed spatially extended sources is less stringent than that based on point-like sources (by a factor varying from an order of magnitude for low mass dark matter particles, to a factor of $\sim$2 for dark matter heavier than $\sim$1 TeV).

\section{Discussion}
\label{discussion}

\begin{figure*}[!]
\mbox{\includegraphics[width=0.49\textwidth,clip]{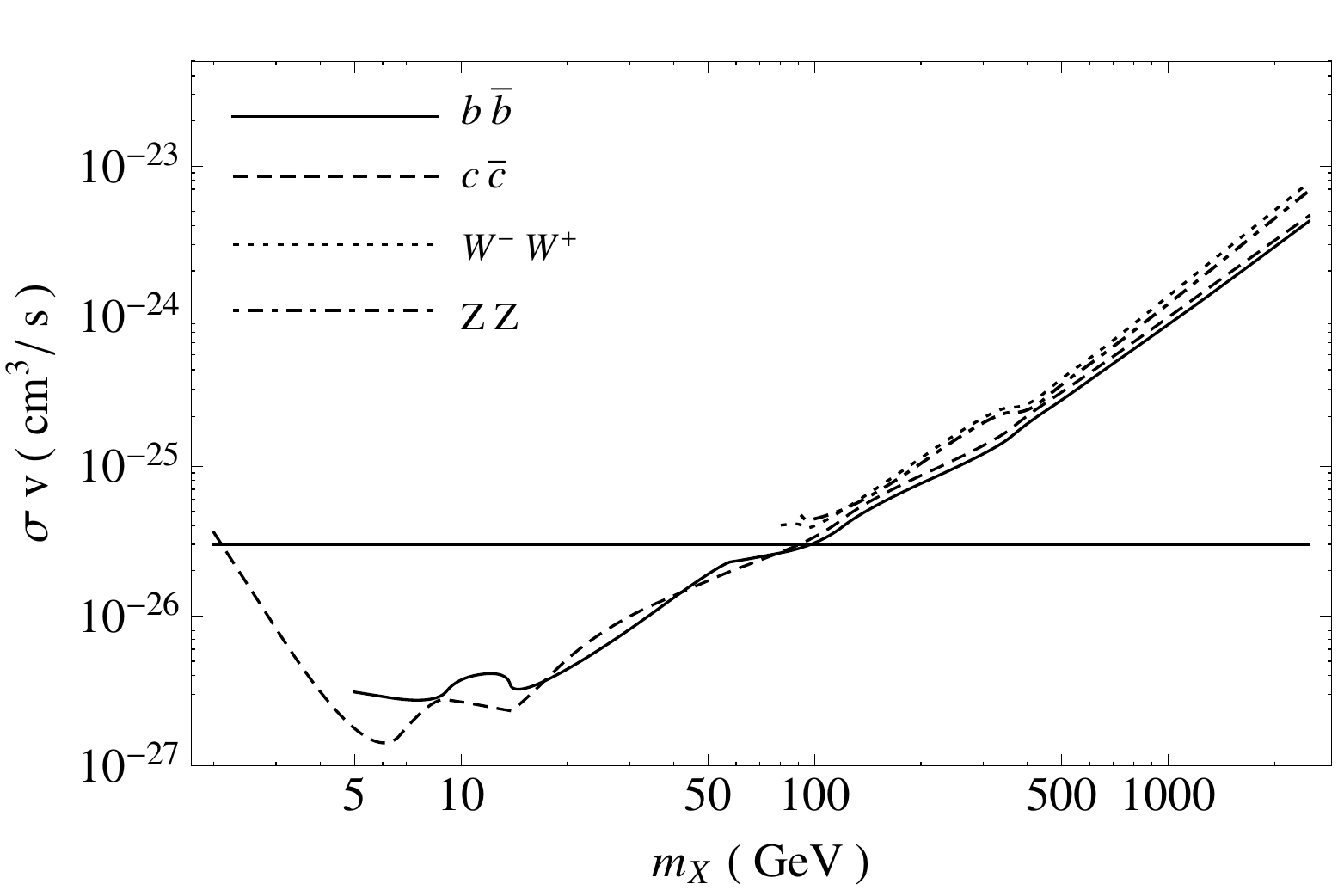}}
\mbox{\includegraphics[width=0.49\textwidth,clip]{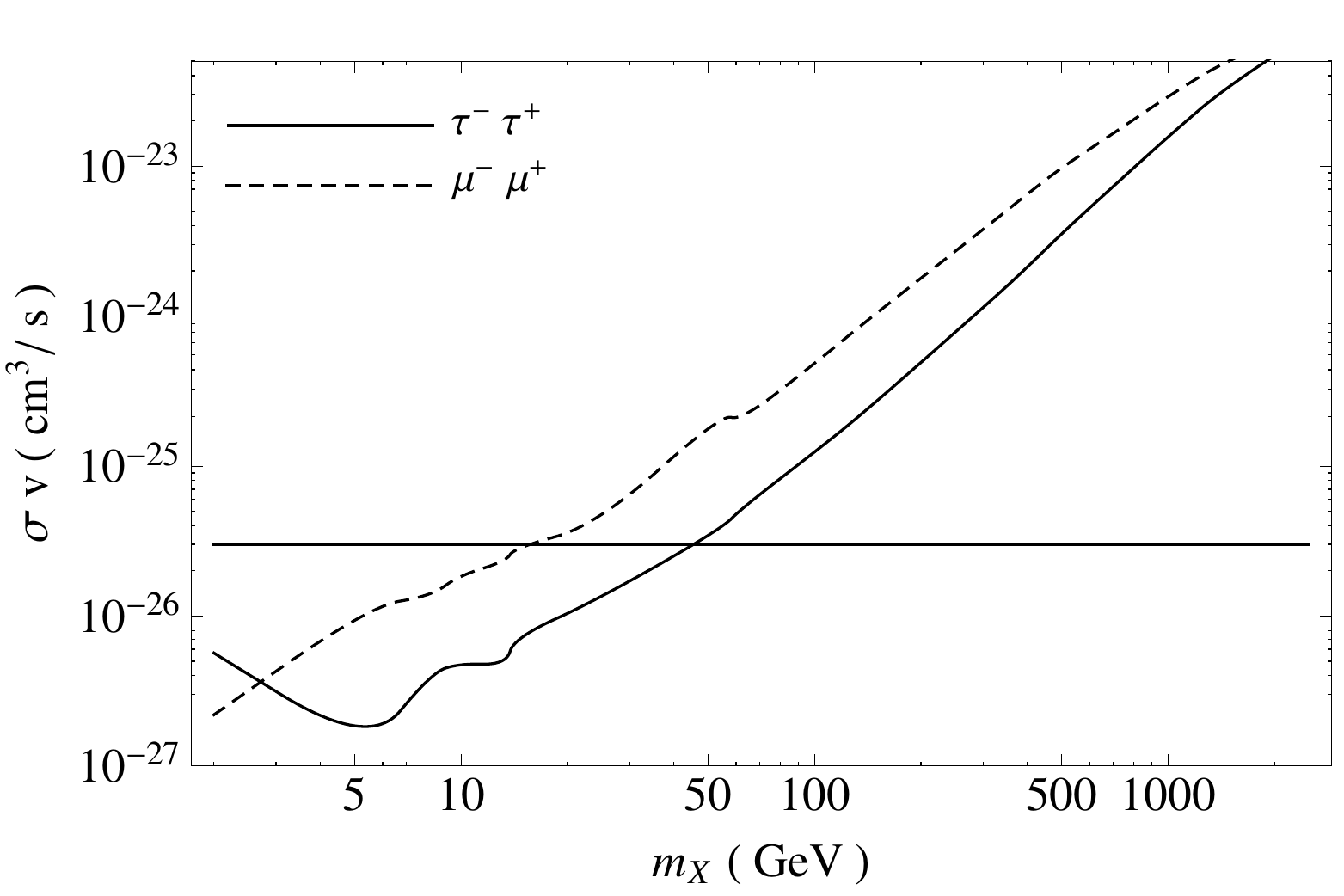}}
\caption{The 95\% confidence level upper limits derived in this study on the dark matter annihilation cross section, for various choices of the dominant annihilation channel. For comparison, the horizontal solid line is the standard estimate for a simple thermal relic ($\sigma v \approx 3 \times 10^{-26} \text{ cm}^3/\text{s}$). For a discussion of related uncertainties, see Sec.~\ref{discussion}.}
\label{limits}
\end{figure*}

The limits derived in this study (and shown in Fig.~\ref{limits}) are quite stringent.  In particular, for dark matter particles with masses below $\sim$200 GeV, they are the strongest presented to date (with the exception of annihilations to $e^+ e^-$ or $\mu^+ \mu^-$, which are currently more strongly constrained by cosmic ray observations~\cite{Bergstrom:2013jra}). For example, while gamma-ray observations from the Milky Way's dwarf galaxies~\cite{GeringerSameth:2011iw,Ackermann:2011wa} and the Galactic Center~\cite{Hooper:2012sr} each rule out dark matter particles with masses below $\sim$30 GeV for an annihilation cross section of $\sigma v=3\times 10^{-26}$ cm$^3$/s to $b\bar{b}$, the limits presented here exclude such dark matter candidates with masses up to $\sim$100 GeV. 

\begin{figure*}[!]
\mbox{\includegraphics[width=0.49\textwidth,clip]{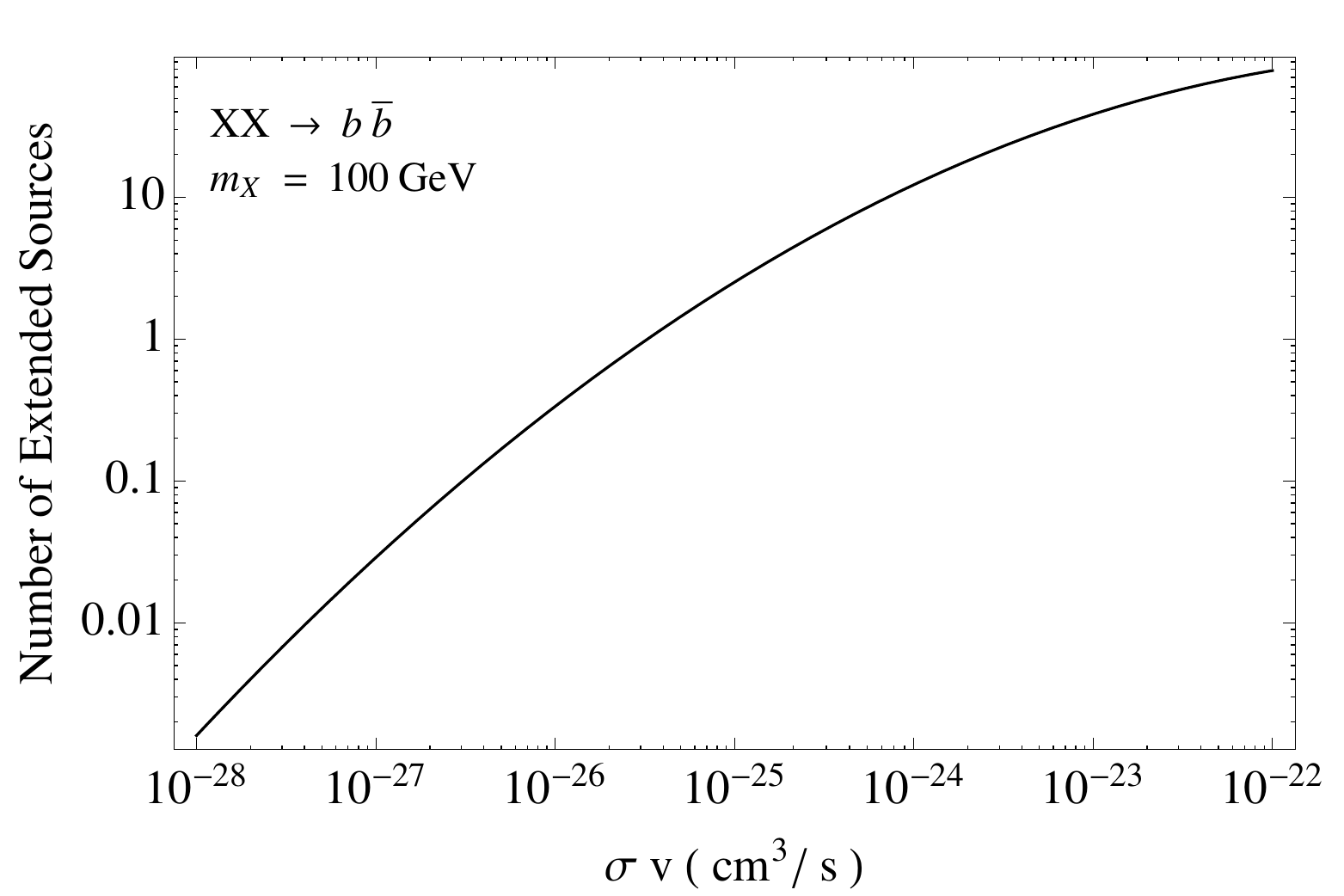}}
\mbox{\includegraphics[width=0.49\textwidth,clip]{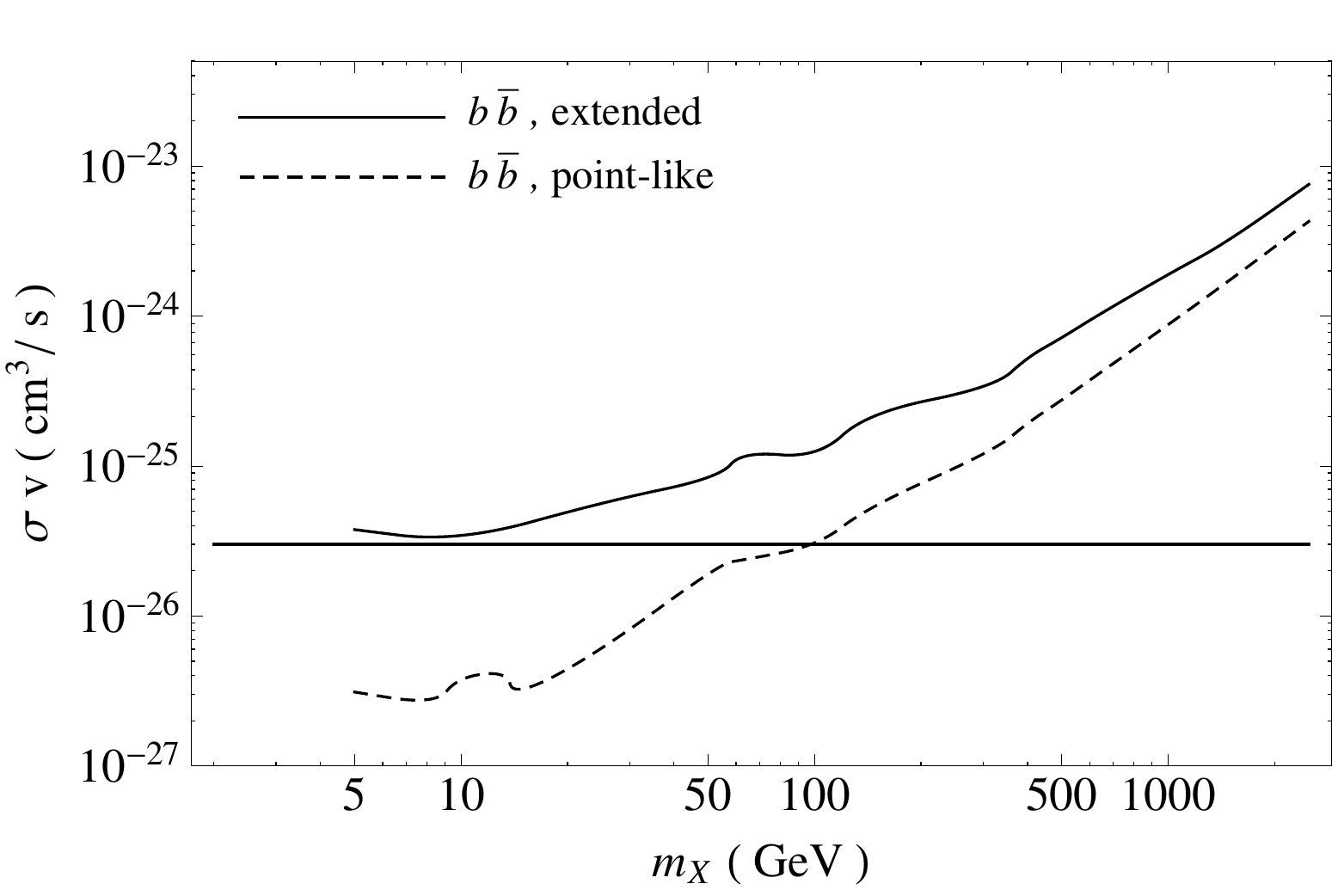}}
\caption{Left: The number of dark matter subhalos predicted to be observed by Fermi with a detectable degree of spatial extension (and located outside of the Galactic Plane, $|b|>10^{\circ}$), as a function of the annihilation cross section, for the case of a 100 GeV particle annihilating to $b\bar{b}$. Right: The 95\% confidence level upper limits on the dark matter annihilation cross section, as derived from the lack of observed spatially extended subhalo candidates (solid), compared to the constraint resulting from the distribution of observed point-like sources (dashed). For all masses, the constraint from the lack of extended sources is weaker than that derived from the point-like source distribution.}
\label{extendedlimits}
\end{figure*}

The limits presented here, however, are subject to uncertainties which, although unlikely to be large, are somewhat difficult to quantify.  For dark matter in the form of cold, collisionless particles (as opposed to warm or strongly self-interacting dark matter), we see no reason that the Milky Way's subhalo population would be significantly different than has been predicted by the Aquarius or Via Lactea simulations.  The Aquarius Project, however, simulated a sample of only six (high-resolution) Milky Way-sized halos, making it difficult to precisely quantify the halo-to-halo variation in the subhalo populations contained within such systems. 
Furthermore, the normalization used in our Eq.~\ref{norm} was based on Aquarius' highest resolution halo (Aq-A), which was also the most concentrated of the six ($c=16.1$ for Aq-A, compared with values ranging from 8.3 to 15.2 for the other five simulated halos). And while the subhalo populations found within each of Aquarius' six simulated halos are quite similar to one another, we cannot entirely discount the possibility that the Milky Way's subhalo population could fall in the tail of the halo-to-halo distribution, and thus differ from these simulated populations. In light of these issues, we acknowledge that such uncertainties could plausibly weaken the limits presented here by a modest factor, perhaps $\sim$2-3, but not much more. As we are unable to quantitatively determine the magnitude of these uncertainties, they are not included in the limits presented in Figs.~\ref{limits} or \ref{extendedlimits}.

There may also be uncertainties introduced as a result of our treatment of tidal stripping. Recall from Sec.~\ref{theory} that we have assumed each subhalo to have lost the outermost 99.5\% of its initial mass due to tidal effects (based on a comparison of the fraction of mass in subhalos locally to that near the Milky Way's virial radius, as seen in Aquarius' halos Aq-1, Aq-2, and Aq-3). This assumption is a fairly conservative one, as some degree of tidal stripping is expected even in the outer volume of the Milky Way's halo, which would imply a larger degree of stripping than the 99.5\% we have assumed, leading to local subhalos with higher densities and greater annihilation rates. Additionally, while we have assumed a common value of 99.5\% mass loss for all nearby subhalos, in reality this is an average quantity, and we expect some subhalos to have lost more or less than this value. Any distribution of mass loss that yields an overall value of 99.5\% across all nearby subhalos, however, leads to a number of observable gamma-ray sources that is at least as large as that predicted under our assumptions.


\clearpage

Lastly, we comment on the possibility that a significant fraction of Fermi's unidentified sources could be the result of annihilating dark matter in nearby subhalos.  Of the ten unidentified, bright ($\Phi_{\gamma} > 10^{-9}$ photons cm$^{-2}$ sec$^{-1}$, between 1-100 GeV), non-variable, high latitude ($|b|>30^{\circ}$) gamma-ray sources observed by Fermi, we note that four of these exhibit quite similar gamma-ray spectra (J2112.5-3042, J1226.0+2953, J1511.8-5013, and J1630.3+3732, see Table~\ref{tab1}). Furthermore, these four sources can each be fit by $\sim$30-60 GeV dark matter particles annihilating to $b\bar{b}$, by $\sim$20-45 GeV dark matter particles annihilating to $c\bar{c}$, or by $\sim$8-10 GeV dark matter particles annihilating to $\tau^+ \tau^-$.  To account for all four of these sources with annihilating dark matter, we require a cross section on the order of $\sigma v \sim (5-10)\times 10^{-27}$ cm$^3$/s (for $b\bar{b}$ or $c\bar{c}$), or $(2.0-2.5)\times 10^{-27}$ cm$^3$/s (for $\tau^+ \tau^-$). These characteristics are intriguingly similar to those required to account for the spatially extended gamma-ray emission observed from the Galactic Center~\cite{gc1,gc2,gc4,newgcchris} and from the surrounding Inner Galaxy~\cite{gc5,Huang:2013pda}.

\section{Summary And Conclusions}
\label{summary}

High-resolution simulations predict that the dark matter halo of the Milky Way should contain a very large number of smaller subhalos. If the dark matter consists of particles with an annihilation cross sections not far below the current upper limits, the largest nearby subhalos could produce a potentially observable flux of gamma-rays. To date, the Fermi Gamma-Ray Space Telescope has detected 82 non-variable, high-latitude ($|b|>30^{\circ}$) gamma-ray sources that have currently not been identified or associated with emission at other wavelengths. In this paper, we have studied these unidentified sources in an attempt to determine whether any might be dark matter subhalos, and have used this information to derive new and quite stringent upper limits on the dark matter annihilation cross section.  

Our most stringent constraints (shown in Fig.~\ref{limits}) were obtained by considering the brightest, high-latitude, unidentified sources detected by Fermi (the ten such sources with $\Phi_{\gamma} > 10^{-9}$ photons cm$^{-2}$ sec$^{-1}$, between 1-100 GeV). For each subhalo candidate source, we performed spectral fits to a variety of dark matter models, and determine which subset of these sources could potentially be members of a dark matter subhalo population. By requiring that the predicted number of subhalos not exceed the number observed with a given spectral shape, we place upper limits on the dark matter's annihilation cross section. For dark matter particles lighter than $\sim$200 GeV, the constraints derived here are the most stringent to date, being modestly stronger than those previously derived from observations of dwarf galaxies or the Galactic Center. 

We also note that four of Fermi's ten brightest, high-latitude, unidentified sources exhibit a similar spectral shape, and therefore represent a particularly promising sample of possible subhalo candidates. This common spectral shape is consistent with arising from $\sim$30-60 GeV dark matter particles annihilating to $b\bar{b}$ with a cross section on the order of $\sigma v \sim (5-10)\times 10^{-27}$ cm$^3$/s, or $\sim$8-10 GeV dark matter particles annihilating to $\tau^+ \tau^-$ with $\sigma v \sim (1-2)\times 10^{-27}$ cm$^3$/s, similar to the characteristics required to account for the spatially extended gamma-ray emission observed from the Galactic Center~\cite{gc1,gc2,gc4,newgcchris} and from the surrounding Inner Galaxy~\cite{gc5,Huang:2013pda}. As Fermi continues to collect data and measure the spectra of these sources with increasing precision, it should become more clear whether they, in fact, have a common spectral shape, and thus could potentially constitute a collection of dark matter subhalos.

\bigskip

{\it Acknowledgements}:  We would like to thank Andrew Hearin, Keith Bechtol, and Tongyan Lin for valuable discussions. We would also like to acknowledge the hospitality of the Aspen Center for Physics, where this work was in part completed. This work has been supported by the US Department of Energy and by the Kavli Institute for Cosmological Physics.

\end{document}